\documentclass[manuscript,screen]{acmart}

\AtBeginDocument{%
  \providecommand\BibTeX{{%
    \normalfont B\kern-0.5em{\scshape i\kern-0.25em b}\kern-0.8em\TeX}}}

\setcopyright{acmcopyright}
\copyrightyear{2018}
\acmYear{2018}
\acmDOI{XXXXXXX.XXXXXXX}

\usepackage{tcolorbox}
\usepackage{multirow}
\usepackage{algorithm}
\usepackage{algorithmic}
\usepackage{hyperref}
\usepackage{indentfirst}
\usepackage{enumerate}

\usepackage[shortlabels]{enumitem}

\usepackage{pifont}
\usepackage{forest}

\newcommand{\circled}[1]{\tikz[baseline=(char.base)]{
    \node[shape=circle, draw=black, fill=black, text=white, inner sep=1pt] (char) {\small\strut #1};}}

\usepackage{fontawesome5}

\begin{document}
\title{Enhancing Android Malware Detection: The Influence of ChatGPT on Decision-centric Task}

\author{Yao Li}
\email{liyao@must.edu.mo}
\affiliation{%
  \institution{Macau University of Science and Technology}
  \city{Macao}
  \country{China}
}

\author{Sen Fang}
\email{fangsen1996@gmail.com}
\affiliation{%
  \institution{North Carolina State University }
  \city{North Carolina}
  \country{USA}
}

\author{Tao Zhang\textsuperscript{*}}
\email{tazhang@must.edu.mo}
\affiliation{%
  \institution{Macau University of Science and Technology}
  \city{Macao}
  \country{China}
}

\author{Haipeng Cai}
\affiliation{%
  \institution{University at Buffalo}
  \city{Buffalo}
  \country{USA}}
\email{haipengc@buffalo.edu}

\renewcommand{\shortauthors}{Y. Li, S. Fang, T. Zhang, H. Cai}
\footnotetext[1]{Tao Zhang\textsuperscript{*} is the corresponding author.}

\begin{abstract}
With the rise of large language models, such as ChatGPT, non-decisional models have been applied to various tasks. Moreover, ChatGPT has drawn attention to the traditional decision-centric task of Android malware detection. 
Despite effective detection methods proposed by scholars, they face low interpretability issues. Specifically, while these methods excel in classifying applications as benign or malicious and can detect malicious behavior, they often fail to provide detailed explanations for the decisions they make. This challenge raises concerns about the reliability of existing detection schemes and questions their true ability to understand complex data.
In this study, we investigate the influence of the non-decisional model, ChatGPT, on the traditional decision-centric task of Android malware detection. We choose three state-of-the-art solutions, $Drebin$, $XM_{AL}$, and $MaMaDroid$, conduct a series of experiments on publicly available datasets, and carry out a comprehensive comparison and analysis.
Our findings indicate that these decision-driven solutions primarily rely on statistical patterns within datasets to make decisions, rather than genuinely understanding the underlying data. In contrast, ChatGPT, as a non-decisional model, excels in providing comprehensive analysis reports, substantially enhancing interpretability.
Furthermore, we conduct surveys among experienced developers. The result highlights developers' preference for ChatGPT, as it offers in-depth insights and enhances efficiency and understanding of challenges. Meanwhile, these studies and analyses offer profound insights, presenting developers with a novel perspective on Android malware detection—enhancing the reliability of detection results from a non-decisional perspective.
\end{abstract}




\maketitle

\setlength{\parindent}{1em}

\setlength{\parindent}{1em}
\section{Introduction}\label{intro}

\noindent The Android operating system is the world's most widely used mobile platform. As of the first quarter of 2024, Android continues to lead the smartphone market, capturing over 70\% of the total market share ~\cite{r1}. However, this widespread adoption has made Android a prime target for malicious attacks. Alarmingly, in 2022, cybercriminals released nearly 135,000 new malware variants per day, equating to over 93 attack attempts per minute, according to G DATA CyberDefense \cite{r2}. Furthermore, Kaspersky blocked almost 33.8 million malware, adware, and riskware attacks in 2023~\cite{r3}. Given Android's popularity, the operating system remains susceptible to a wide array of attacks, including credential theft, privacy breaches, bank fraud, ransomware, adware, SMS spoofing, and more. The Android malware threat poses a significant danger to users and continues to escalate as malware becomes increasingly sophisticated and virulent. For instance, ransomware like \textit{LockerPin} can lock users out of their devices, while spyware such as \textit{Pegasus} can steal sensitive data without detection. Additionally, banking trojans like \textit{}EventBot can siphon financial information, further highlighting the growing severity of this issue. Consequently, the accurate classification of malware applications is paramount to ensuring system security and safeguarding user privacy~\cite{r4}.

Numerous approaches have been proposed to detect Android malware \cite{r28, r74, Cai, XuDroid,r80,r83}, aiming to uncover new attack patterns, devise fresh signatures, or identify malicious code. For instance, Feng et al. \cite{r7} propose semantic-based Android malware detection through static analysis. Cai et al. \cite{r24} propose a series of weighted formulas to improve the detection effect of models such as support vector machine (SVM). Kim et al. \cite{r31} propose a multi-modal deep learning (DL) model to detect Android malware. Meanwhile, in Android malware detection, beyond basic identification, there is an emphasis on categorizing malware into specific classes for improved recognition \cite{r63,chakraborty_pierazzi_subrahmanian_2020,elish_elish_almohri_2022}. For instance, Xu et al. \cite{r60} utilize deep learning to autonomously segment Android malware into distinct categories. Similarly, Vij et al. \cite{r62} introduce a method hinged on a graph signature-based classification approach. However, whether these methods are based on traditional machine learning or advanced deep learning techniques, whether their goal is detection or classification, they all face low interpretability issues. While some solutions, such as Drebin \cite{r19}, claim to offer explainable malware detection, they often fall short in providing the level of detailed analysis and explanations that can fully inform developers about the basis of their decision-making. This has raised concerns about the reliability of existing detection schemes. Especially when these existing solutions are deployed in real-world data or when new datasets are introduced, the results for detecting new unknown malicious applications tend to be subpar. This decline in detection performance further exacerbates our concerns about their decision-making reliability. Moreover, it also leads developers to doubt the true capability of these solutions in understanding complex Android applications.

Simultaneously, with the birth and evolution of large language models like ChatGPT, their excellent analysis and understanding capabilities are gradually exploited. This powerful ability to analyze and explain has led to ChatGPT being employed across various tasks \cite{gpt4}. With appropriate prompts, ChatGPT can excel in diverse tasks, meticulously analyzing data and providing detailed explanations. The limitations of current solutions, combined with ChatGPT's analytical and capabilities, have led us to reconsider the following question: \textbf{Under the influence of ChatGPT, is there a novel perspective that detects Android malware from a non-decisional viewpoint, emphasizing not just the act of deciding, but also explaining the reason behind those decisions?}

To answer this question, we undertake the first comprehensive empirical study, delving into the limitations of decision-based models in Android malware detection tasks and the influence of ChatGPT on existing Android malware detection tasks. One of the key reasons for choosing ChatGPT over other large language models (LLMs) is that it requires no additional training. This significantly lowers the technical barriers and reduces the time and resources typically needed to fine-tune or retrain models. Unlike some other LLMs that may require customization or fine-tuning for specific tasks, ChatGPT is a ready-to-use model, providing high-quality results out of the box. Additionally, it eliminates the need for extensive computational resources, which are often required for training large models, making it an efficient and cost-effective solution. Meanwhile, we select three state-of-the-art detection solutions: $Drebin$, $XM_{AL}$, and $MaMaDroid$. Among them, $Drebin$ and $XM_{AL}$ are based on a machine learning model while $MaMaDroid$ is built upon a deep learning model. We then craft appropriate prompts to guide ChatGPT in Android malware detection tasks. We input the same dataset (features extracted from both benign and malware) into $Drebin$, $XM_{AL}$, $MaMaDroid$, and $ChatGPT$ respectively, retrieve their outputs, and subsequently conduct a comparative analysis. To structure our investigation, we formulate three research questions (RQ) to validate our observations. We provide a detailed exposition in the subsequent subsections.

\subsection{Reliability Analysis}
\noindent To gain a deeper understanding of Android malware detection and ChatGPT, we conduct an in-depth study of three state-of-the-art methods: $Drebin$ and $XM_{AL}$ based on machine learning, and $MaMaDroid$ built upon deep learning. Our analysis is guided by the following research questions:

\begin{tcolorbox}[title = {RQ 1:},
  colframe = gray!30!white, colback = gray!10!white,
  colbacktitle = gray!30!white,
  coltext = black!50!black,
  coltitle = black!90!white]
  To what extent can current models proficiently identify Android malware, and how reliable are they in terms of comprehending data when compared to ChatGPT?
\end{tcolorbox}

We conduct a series of experiments employing $Drebin$, $XM_{AL}$, $MaMaDroid$, and ChatGPT on the same dataset. Our research findings indicate that existing solutions can effectively detect malicious software, but they are susceptible to dataset biases and lack interpretability. \textbf{In contrast, ChatGPT, while unable to provide specific decisions, can offer comprehensive and detailed analysis. It categorizes the input, analyzes each category, and ultimately provides a comprehensive analysis of potential risks and a maliciousness score.} For more information, please refer to Section \ref{relia}.

\subsection{Influence Analysis}

\noindent Following ChatGPT's analysis of each sample, we obtain a series of detailed reports. We meticulously scrutinize both ChatGPT's reports and the outputs generated by existing models for each sample.

\begin{tcolorbox}[title = {RQ 2:},
  colframe = gray!30!white, colback = gray!10!white,
  colbacktitle = gray!30!white,
  coltext = black!50!black,
  coltitle = black!90!white]
  How does the large language model (i.e., ChatGPT) impact existing Android malware detection solutions? And what valuable insights can it provide to motivate developers?
\end{tcolorbox}

Through comprehensive analysis and extensive research, we identify that existing solutions suffer from usability issues, being not only challenging to use but also presenting difficulties in terms of re-deployment and reproducibility for programmers. Additionally, the low interpretability of current solutions is a pressing concern. In contrast, ChatGPT offers comprehensive and detailed analysis but falls short in decision-making capabilities. \textbf{This realization leads us to understand that the development of Android malware solutions should not be solely driven by decision-centric tasks but should emphasize the need for a more interpretable approach that provides explanations alongside decision-making.} For more details, please refer to Section \ref{impact}.

\subsection{Improvement Analysis}
\noindent Dataset bias and low interpretability are prevalent concerns within existing Android malware solutions. It prompts questions about these models' capacity to genuinely comprehend the data. Thus, in this study, we strive to enhance and improve Android malware detection through a novel perspective.

\begin{tcolorbox}[title = {RQ 3:},
  colframe = gray!30!white, colback = gray!10!white,
  colbacktitle = gray!30!white,
  coltext = black!50!black,
  coltitle = black!90!white]
  How to enhance the Android Malware Detection capability of existing large language models (i.e., ChatGPT)?
\end{tcolorbox}

Through our research, we discover that existing Android malware detection solutions, while effective, still suffer from issues related to dataset bias and low interpretability. Meanwhile, although ChatGPT can provide detailed analysis and explanations, it cannot make final decisions. Based on these findings, we aim to provide developers with more insights from a non-decision perspective to assist them in making improvements in the future. \textbf{Therefore, we find two ways to improve: 1) Further enhancing the explanatory and analytical capabilities of ChatGPT; 2) Constructing a large model specifically for detecting Android software.} For the details, please refer to Section \ref{improve}.

\subsection{Contributions}
\noindent In summary, this paper makes the following contributions:

\begin{itemize}
\item We conduct the \textbf{first comprehensive study} on the impact of ChatGPT in Android malware detection, identifying its strengths in interpretability and analysis compared to existing solutions.
\item Through experiments and user/developer surveys, we demonstrate the strong preference for large language models like ChatGPT due to their interpretability and analytical power.
\item We introduce a novel perspective for future Android malware detection: prioritizing \textbf{explanation alongside decision-making} to enhance usability and trust in large language models.
\end{itemize}

The remainder of this article is structured as follows. Section \ref{backgrd} introduces the motivation and the relevant information on ChatGPT and Android malware detection. The proposed approach is presented in Section \ref{method}. Section \ref{setup} describes the experimental setup and evaluation metrics. Experimental results are shown in Section \ref{study}. The further analysis of the proposed approach is discussed in Section \ref{discuss}. Section \ref{R-work} presents the related work. Finally, Section \ref{coclusion} provides a summary of the proposed approach in this paper.

{\setlength{\parindent}{0cm}
\textbf{Data Avalibility:} The data underlying this paper is available on Figshare \footnote{\url{https://doi.org/10.6084/m9.figshare.27004879.v1}}.

\setlength{\parindent}{1em}
\section{Motivation and Background}\label{backgrd}
\subsection{Motivation}

\noindent The development of Android malware detection has progressed significantly over the years, giving rise to numerous effective detection strategies. In theory, these strategies have demonstrated remarkable efficacy. However, they still suffer from dataset bias, particularly selection bias and label bias, which stem from imbalanced and inconsistent datasets used during model training. To investigate this, we select six representative strategies for our study and conduct experiments to evaluate their performance. 
\textbf{Therefore, we collected 67 instances of new malware introduced between July 1, 2023, and August 31, 2023, from the MalwareBazaar\footnote{\url{https://bazaar.abuse.ch/}} website. This platform uploads newly discovered malware daily, providing direct download access to researchers and analysts.
} The results of their detection are presented in Table \ref{z-day}.

\begin{table}[h]
\centering
\caption{Performance of schemes against Zero-day Malware.}
\label{z-day}
\begin{tabular}{ccccccc}
\hline
Method         & EFIMDetector & Drebin & MaMaDroid  & EC2   & SEDMDroid & MMN  \\ \hline
Number         & 22           & 9      & 33             & 26    & 22        & 26          \\
Detection Rate & 0.343        & 0.143  & 0.497       & 0.401 & 0.339     & 0.389     \\ \hline
\end{tabular}
\end{table}

From Table \ref{z-day}, it can be observed that the six schemes perform poorly in detecting unknown malicious applications. While these methods have demonstrated good results in training and testing datasets, a significant decline can be observed when detecting unfamiliar malware samples. This phenomenon raises our concerns regarding the reliability of the detection results. Meanwhile, these detection solutions have not provided detailed analysis and explanations. This means that we do not know the basis of their decision-making. Moreover, this raises our concerns about the reliability of these detection schemes. Consequently, we undertake a more in-depth exploration. As illustrated in Figure \ref{fig:m1}, we display the detection report of unknown application on the MalwareBazaar\footnote{\url{https://bazaar.abuse.ch/sample/44d0a81bdd9bbd3892f0ec49c430d60717be35e15a1a9ff3c2b4fddae58ad45c/}} website. This site uploaded the application to eight popular antivirus tools. Among them, two tools deem the software harmless, but three label this unknown software as malicious, two consider it suspicious, and one does not provide any result. This indicates that even these advanced antivirus tools are unable to make a unanimous decision regarding an unknown malicious application. All of these tools provide their detection results, with some even offering detailed reports. However, these reports suffer from poor readability and require a wealth of specialized knowledge to comprehend. Moreover, these reports do not provide the rationale behind their decisions, nor do they offer explanations or analyses. Therefore, this demonstrates that both advanced detection schemes and popular antivirus tools have neglected the importance of explanation and analysis. This situation motivates us to conduct in-depth research to improve interpretability, thereby enhancing Android malware detection.

{Meanwhile, we conducted a survey involving 101 practitioners to gather insights into their perspectives on current Android malware detection methods. Each participant was asked six questions, as listed in Table \ref{tab:survey_questions_on_chatgpt}. The survey was designed in a semi-close-ended format, allowing participants to provide additional written feedback to supplement their answers. 

{The participants held diverse roles, including software engineers, developers, managers, and other professionals. Among them, 13 individuals were directly involved in the development of antivirus software, offering specialized insights into the field. The remaining participants worked in related areas such as mobile software development, operations, and maintenance. Table \ref{tab:demography_of_survey_participants} provides a detailed breakdown of the participants' professions and years of experience.

{To select participants, we employed the snowball sampling approach~\cite{Goodman-SnowballSampling-AnnMathStat1961}. Initially, we reached out to 19 senior professionals from our personal contacts, each with over 10 years of industry experience. After recording their feedback, we asked them to refer other suitable candidates for the study. This process led to the inclusion of an additional 82 participants, encompassing a broad spectrum of roles, from developers to managers. This approach ensured a diverse and experienced pool of respondents for the survey. The following is a detailed analysis of the questionnaire survey results.

\circled{Q1} In response to the first question, "Do you use antivirus software?", all respondents reported utilizing antivirus software to protect their devices. This widespread adoption highlights the critical role of antivirus solutions as an essential tool for ensuring security, both for individuals and businesses. Whether for personal or professional use, antivirus software enjoys considerable popularity, with individuals generally perceiving it as a crucial measure for ensuring system security. This demonstrates that respondents, regardless of their technical expertise, exhibit a notable degree of protection awareness.
 
\circled{Q2} Second question, we ask the participants about their reaction/experience of using antivirus software. The vast majority of respondents (91\%) express satisfaction with the performance of their antivirus software, indicating that the majority of users believe it can effectively protect their systems. However, a notable minority (5\%) report negative experiences, suggesting some level of dissatisfaction or a neutral attitude. Additionally, 4\% of respondents indicate that their experience with this software is not positive and identify potential areas for improvement, particularly in functional performance and user experience.
 
\circled{Q3} The third question posed to the interviewees is: "How well do you understand the detection reports generated by antivirus software?" With regard to the comprehension of the detection reports, the distribution of respondents' answers is relatively dispersed. Approximately 34.7\% of respondents indicate that they could fully comprehend the report content, suggesting that these individuals possess a robust technical background or familiarity with the report's subject matter. However, 11.9\% of respondents indicate that their understanding of the report is limited, suggesting that the technical terminology or structure of the report may present a challenge. Of greater concern is that more than half of the respondents (53.4\%) do not understand or are unclear about the report content, indicating that antivirus software reports may be too complex for ordinary users and lack simple and clear explanations.
 
\circled{Q4} The survey results of the fourth question indicate that 95\% of respondents had utilized ChatGPT, which is indicative of the pervasive adoption of this AI tool across technical and non-technical domains. This finding underscores the recognition of ChatGPT as a pivotal problem-solving instrument and its significant role in daily work activities. Conversely, only 2.9\% of respondents report no usage of ChatGPT, which could be attributed to their professional requirements or a lack of familiarity with such tools.
 
\circled{Q5} The fifth question is about the performance of ChatGPT. To ensure the interpretability of ChatGPT in SE real-world situations, we need to assess its ability to- i) interpret specific scenarios and contexts accurately; ii) generate appropriate information based on that. Hence, we provide ChatGPT with a specific question-answer set (about \texttt{Glide}) from Stack Overflow and ask it to explain how the library is used solely based on the provided information. In response to the question of whether ChatGPT could provide a more detailed explanation of the problem, 97\% of respondents indicate that they find the explanation provided by ChatGPT to be sufficiently detailed and clear. This suggests that this cohort of users is content with their capacity to resolve issues. Nevertheless, 3\% of respondents indicate a preference for more comprehensive explanations from ChatGPT, suggesting a need for more detailed information.
 
\circled{Q6} The last question posed to the interviewees is: "What can be the ways to enhance existing forms of malware detection?" In regard to the enhancement of malware detection, 69.3\% of respondents indicate that the introduction of AI and machine learning technology would facilitate the improvement of detection capabilities. This finding suggests a high level of expectation for the integration of automation and intelligent technology in this domain. A 9.9\% proportion of respondents express the desire to improve both the detection function and the software's accuracy in identifying threats, indicating that existing functions could benefit from enhancement. Furthermore, 3\% of respondents underscore the necessity for simplifying the user experience, proposing that a more intuitive interface and operational methodology would facilitate the software's usability. The remaining 17.8\% of respondents provide responses that are unclear, indicating that they may lack specific suggestions on how to improve malware detection or may be uncertain about their specific needs.

By analyzing the above-mentioned investigation results, we can get the following conclusion. While existing malware detection schemes are effective, there is still considerable room for improvement in understanding reports and further improving software functionality. In particular, there is a need to simplify the user experience and increase the clarity of reports. Therefore, the survey results motivate us to strengthen the existing detection form.

\begin{table*}
\caption{Survey questions.}
\label{tab:survey_questions_on_chatgpt}
\begin{tabular}{c l}
\hline
\textbf{Q\#} & \textbf{Questions}                                                                                          \\ \hline
1            & Do you use antivirus software?                                                                                                   \\

2            & How would you describe your experience with using it so far?                                                                  \\
3            & How much do you understand the antivirus software's detection reports?                                                          \\
4            & Do you use ChatGPT?                            \\
5            & Do you think ChatGPT can explain in more detail when using ChatGPT to solve problems?                                       \\
6           & What can be the ways to enhance existing forms of malware detection?                                                     \\ \hline
\end{tabular}%
\end{table*}

\begin{table}[]
\centering
\caption{Demography of Survey Participants}
\label{tab:demography_of_survey_participants}
\begin{tabular}{rcccccc}
\textbf{}                                                                                          & \multicolumn{6}{c}{Years of Experience}                                \\ \cline{2-7} 
\multicolumn{1}{r|}{Current Profession}                                                            & 0-5 & 6-10 & 11-15 & 16-20 & \multicolumn{1}{c|}{21-24} & Total        \\ \hline
\multicolumn{1}{r|}{\begin{tabular}[c]{@{}r@{}}Software Engineer\\ (Dev/QA/Data/Ops)\end{tabular}} & 40  & 17   & 11    & 8     & \multicolumn{1}{c|}{-}     & 76 \\ \hline
\multicolumn{1}{r|}{\begin{tabular}[c]{@{}r@{}}Manager\\ (Dev/PM/Ops)\end{tabular}}                & 3   & 1    & 6     & 2     & \multicolumn{1}{c|}{-}     & 12           \\ \hline
\multicolumn{1}{r|}{\begin{tabular}[c]{@{}r@{}}Executive\\ (Company Leadership)\end{tabular}}      & 0   & -    & -     & 1     & \multicolumn{1}{c|}{1}     & 2            \\ \hline
\multicolumn{1}{r|}{\begin{tabular}[c]{@{}r@{}}Non-Tech\\ (Sales/Marketing/CS/HR)\end{tabular}}    & 7   & 3    & 1     & -     & \multicolumn{1}{c|}{-}     & 11           \\ \hline
\multicolumn{1}{r|}{Total}                                                                         & 50  & 21   & 18    & 11    & \multicolumn{1}{c|}{1}     & 101         
\end{tabular}
\end{table}

\begin{figure}[htb]
\centering
\includegraphics[width=0.9\textwidth]{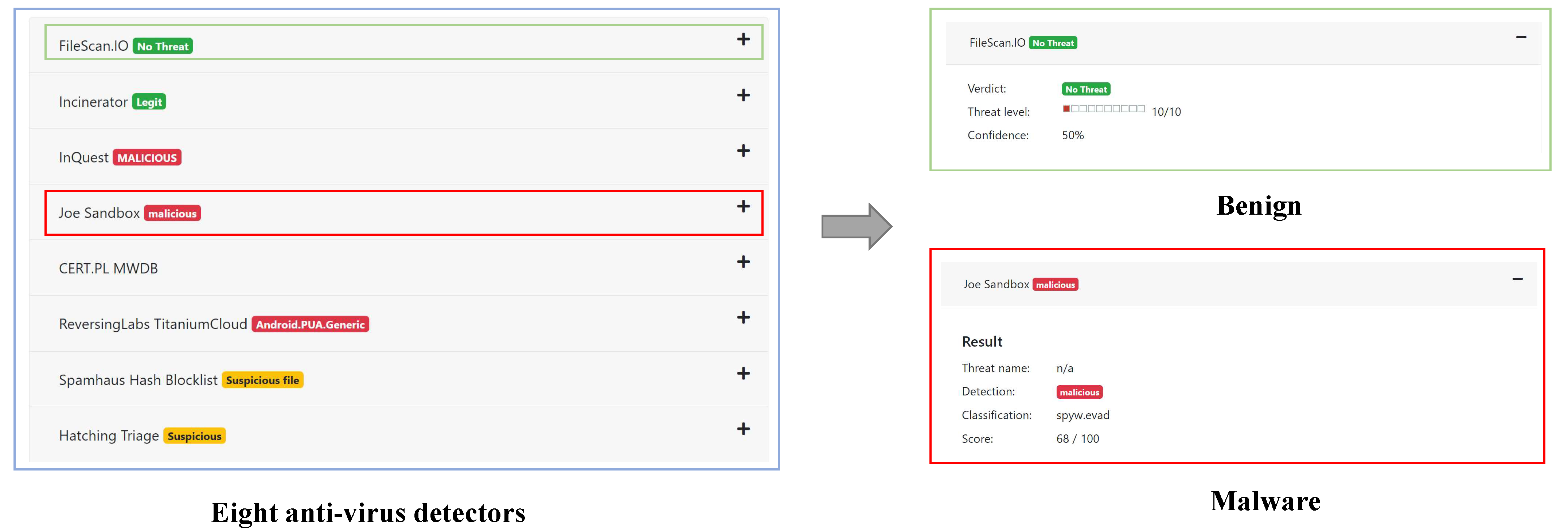}
\caption{\label{fig:m1} SHA256 for "feb1b1019a31d0677c1d25bb80549a171c5f7da381f38014aae0d63b56126722", an unidentified Android software reported on MalwareBazaar.}
\end{figure}

\subsection{ChatGPT}
\noindent ChatGPT originates from OpenAI and is a descendant of the GPT model lineage. From GPT-1 to GPT-3 \cite{gpt1,gpt2,gpt3}, each iteration has seen these models grow exponentially in complexity and computational capacity. ChatGPT adopts a transformer architecture, specifically tailored for sequential data processing, making it particularly suited for text. This structure aids the model in understanding context, thereby generating intricate and coherent text. The versatility of ChatGPT is evident in its wide range of applications. Whether it is content creation, tutoring, coding assistance, language translation, or casual conversation, this model can satisfy diverse conversational needs.

Despite ChatGPT’s powerful capabilities, it is not without challenges. OpenAI prioritizes user safety and the ethical behavior of the model \cite{gpt4}. The fine-tuning phase is supported by human reviewers and guided by clear protocols, aiming to curb unsolicited outputs. OpenAI actively seeks feedback, using insights to refine and enhance model behavior.

In conclusion, ChatGPT has become a beacon in conversational AI, embodying a synthesis of extensive knowledge and vital adaptability to numerous text-generation tasks.

\subsection{Android malware detection}
\noindent In an era where smartphones are omnipresent and the Android operating system is extensively used, the convenience offered by mobile applications is being undermined by the continuous threats posed by the evolution of Android malware. Given the indispensable role that these devices play in our lives, they have also become the primary targets for malicious actors aiming to exploit vulnerabilities and compromise security. The surge in Android malware has stimulated the development of sophisticated detection technologies and tools. Android malware detection can be categorized into three methods: static analysis \cite{r34,r35}, dynamic analysis \cite{r46,r48}, and hybrid analysis \cite{r54,r84}. Static analysis involves decompiling an Android application package (APK) through reverse engineering to extract features from a series of files. Dynamic analysis, on the other hand, involves installing APKs on simulators or real devices to monitor system calls. Hybrid analysis employs both static and dynamic analysis methods. Furthermore, with the evolving landscape of Android malware, there is a need for detection solutions to be continuously enhanced and strengthened.

\section{METHODOLOGY}\label{method}
\begin{figure}[htb]
\centering
\includegraphics[width=0.95\textwidth]{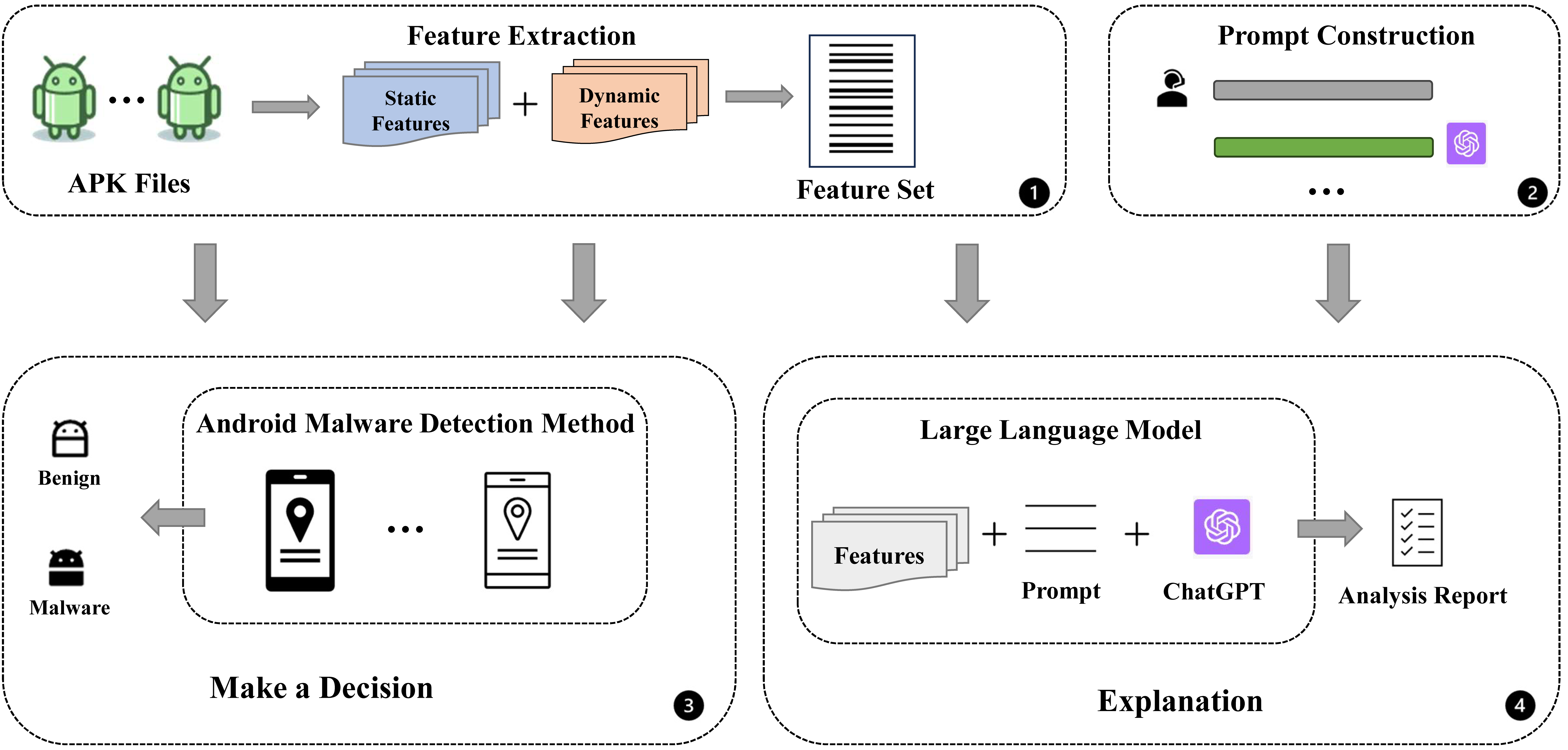}
\caption{\label{fig:over}Overview of our study.}
\end{figure}
\subsection{Overview}

\noindent To delve deeply into the challenges of limited data comprehension, reduced reliability, and interpretability in prevailing Android malware detection solutions, as well as to assess the potential influence of ChatGPT on these solutions, a methodological approach is essential. Our aim is to shed light on their capability to discern features effectively. In this research, as visualized in Figure \ref{fig:over}, we introduce a verification method designed to rigorously assess the dependability of judgments rendered by current Android malware detection strategies.

Initially, we harness reverse engineering methods to distill static features from APK files. Concurrently, to glean dynamic features, the APK is installed and exercised on a genuine Android device.

Next, we craft targeted prompts to steer ChatGPT meticulously, honing its attention on Android malware detection, culminating in the formulation of a conclusive prompt.

The extracted data is then channeled into the prevailing Android malware detection systems and ChatGPT for evaluation.

In the final step, we collate in-depth analytical reports generated by ChatGPT, juxtaposed with individual detection outcomes produced by the conventional detection models.

\subsection{Feature Extraction}
\noindent
The APK file, which functions as the installation package for Android applications, is analyzed using reverse engineering tools such as Apktool \cite{rTool} and Androguard \cite{rGuard}. Apktool decompiles the APK into its core resources and bytecode, extracting critical components like the \textit{AndroidManifest.xml} file and converting the \textit{Class.dex} file (containing Dalvik Executable bytecode) into a human-readable \textit{.smali} format. This enables detailed examination of API calls, strings, and control flows. In contrast, Androguard offers a robust framework for static analysis, supporting disassembly, decompilation, and in-depth bytecode inspection. It complements Apktool by facilitating advanced operations such as cryptographic signature analysis and comprehensive behavioral insights.

The extracted \textit{AndroidManifest.xml} file is parsed to collect static features such as declared permissions, intent filters, and metadata, providing valuable information about the app's functionalities and interactions with the Android system. Additionally, the \textit{Class.dex} file, once converted to \textit{.smali} format, supports an in-depth exploration of the app's logic, helping uncover potential malicious behaviors embedded within its control flows or API usage patterns.

For dynamic feature extraction, applications are executed in a controlled emulator environment where runtime behaviors are monitored. This includes tracking system calls, interactions with the Android system, and other runtime characteristics. 

When obfuscated APK files are encountered, unpacking tools \cite{r66, r67} are employed to recover the underlying files. Features extracted from these samples are cross-verified against dataset-provided features to ensure their accuracy.

\subsection{Prompt Construction}\label{prompt}
\noindent A prompt acts as a director to initiate a machine learning model, particularly those designed for text generation. It is essentially a text fragment or directive that shapes the model's output towards a specific theme, style, or structure. Unlike the conventions of supervised training, where every sample needs an associated label, using prompts in training only necessitates a guiding statement to steer the model's output generation. As such, the prompt's formulation becomes paramount. Inappropriate or vague prompts could lead the model astray, yielding incorrect or irrelevant results. We have delineated a three-step process to craft an effective prompt:
\begin{enumerate}
    \item \textbf{Task Definition:} Starting by succinctly expressing the specific output the prompt aims to elicit from the model.
    \item \textbf{Role Assignment:} Ascertaining and detailing the role the model should embody during text generation. Setting a clear context or perspective for the model often refines the resultant output, aligning it closely with the intended purpose.
    \item \textbf{Guidance:} Offering comprehensive directions for the model's text composition. This can encompass specifying the nature of the content (e.g., Answers should reflect advanced mathematical understanding), the structure of the response (this includes aspects like length, format, and tone), and any recommended approaches or strategies for formulation.
\end{enumerate}    

Such a structured approach ensures that the model not only produces relevant content but also maintains a high caliber of output quality. Therefore, we follow the above three principles and build a suitable prompt.
\vspace{-0.2cm}
\tcbset{colback = white}
\begin{tcolorbox}
You are an experienced Android developer and Android malware detector. I will give you the permissions, APIs, etc. used in the APK. Can you analyze these contents and make a judgment as to whether there are problems with this APK, based on these characteristics give the software a malicious score (range 0-1).
\end{tcolorbox}
\vspace{-0.3cm}

\section{Experimental setup}\label{setup}
\noindent To comprehensively assess the reliability of current models for data comprehension, we conducted a series of experiments using publicly accessible datasets. Specifically, we utilized the Kronodroid dataset \cite{Kronodroid_Guerra}, which includes data collected from both emulator environments and actual devices. The dataset comprises 28,745 malicious samples across 209 distinct malware families and 35,256 benign samples. For our analysis, we focus on extracting 289 dynamic features (e.g., system calls) and 200 static features (e.g., permissions, intent filters, and metadata). In our research, we randomly selected 100 benign samples and 100 malicious samples from the Kronodroid dataset, enabling rapid experimentation and analysis. Additionally, we constructed \textbf{$dataset_T$} by randomly selecting 3,000 benign samples and 3,000 malicious samples from the Drebin \cite{r19} and AMD datasets \cite{r11}. Unlike prior studies that emphasize malware family classification, our research exclusively focuses on the binary classification of benign versus malicious applications, disregarding family information throughout dataset construction, model training, and detection processes. For obfuscated samples, we employed unpacking systems \cite{r66, r67} to recover original files, verified the extracted features against dataset-provided features, and ensured the accuracy of the analysis. 

We construct two distinct datasets to support our analysis. The \textbf{$dataset_A$}, containing 200 samples (100 benign and 100 malicious), is primarily used to analyze ChatGPT's performance and conduct comparative experiments. We adopt this size based on the dataset configurations commonly used in related research \cite{ALIGOMBE2018235, Abraham2015}. Its main purpose is to explore ChatGPT's ability to analyze malware rather than optimizing detection performance, and its size also helps minimize computational overhead, fitting within GPT-4's usage constraints. The \textbf{$dataset_D$}, with 6,000 samples (3,000 benign and 3,000 malicious), is used for training traditional detection models, reflecting typical configurations in related research \cite{r28}. The \textbf{$dataset_D$} allows for better model training and generalization by covering a wider range of software behaviors. By using both datasets, we leverage the strengths of each: the smaller one enables quick analysis, while the larger one supports more comprehensive training and validation. The \textbf{$dataset_A$} is utilized for the analysis and testing of ChatGPT, while \textbf{$dataset_D$} is employed for the training and testing of traditional models. For a comprehensive overview of the experiments and results, please refer to Section \ref{relia}.

The subsequent experiments are conducted on a personal computer integrated with 13th Gen Intel(R) Core(TM) i9-13900K 3.00 GHz and NVIDIA GeForce RTX 4090. The computer has 32GB of memory and 2TB of storage. The deep neural networks are implemented using Scikit-learn, and Pytorch. Meanwhile, the ChatGPT 4 model is employed for analysis in subsequent experiments.

\subsection{Evaluation Metrics}
\noindent There are four common metrics adopted to evaluate the performance of classification: accuracy, precision, recall rate, and F-score. The malicious sample is denoted as the positive (P) class and the benign sample is denoted as the negative (N) class. Then, four numbers are obtained:
\begin{itemize}
\item True Positive (TP): the number of positive samples that are
correctly predicted as positive.
\item False Negative (FN): the number of positive samples that
are incorrectly predicted as negative.
\item False Positive (FP): the number of negative samples that are
incorrectly predicted as positive.
\item True Negative (TN): the number of negative samples that
are correctly predicted as negative.
\end{itemize}

Based on these designations, the following equations calculate four common metrics:
\begin{equation}
\text { Accuracy }=\frac{\mathrm{TP}+\mathrm{TN}}{\mathrm{TP}+\mathrm{FN}+\mathrm{FP}+\mathrm{TN}} \\
\end{equation}
\begin{equation}
\text { Precision }=\frac{\mathrm{TP}}{\mathrm{TP}+\mathrm{FP}} \\
\end{equation}
\begin{equation}
\text { Recall }=\frac{\mathrm{TP}}{\mathrm{TP}+\mathrm{FN}} \\
\end{equation}
\begin{equation}
F \text {-Score }=\frac{2 \times \text { Precision } \times \text { Recall }}{\text { Precision }+ \text { Recall }}
\end{equation}

\section{EMPIRICAL STUDY}\label{study}
\noindent In this study, we explore the influence of ChatGPT on decision-making scenarios such as Android malware. Like the recent series of conversational large language models, especially those represented by ChatGPT, they have a higher degree of understanding of data and tasks. By comparing ChatGPT with the traditional decision-making Android malware detection model, we discover the influence of ChatGPT on the traditional model, and how to further improve the traditional model and construct a large language model more suitable for Android malware detection. Therefore, we design three research questions to conduct a comprehensive analysis and research, which are presented in the following subsections \ref{relia}-\ref{improve}.

\subsection{Reliability of data understanding}\label{relia}
\begin{tcolorbox}[title = {RQ 1:},
  colframe = gray!30!white, colback = gray!10!white,
  colbacktitle = gray!30!white,
  coltext = black!50!black,
  coltitle = black!90!white]
  To what extent can current models proficiently identify Android malware, and how reliable are they in terms of comprehending data when compared to ChatGPT?
\end{tcolorbox}

\noindent In this section, we begin by verifying the effectiveness of existing detection solutions and identifying their limitations (Section \ref{per_m}). We conduct this experiment using two datasets: \textbf{$dataset_T$} and \textbf{$dataset_A$}. The \textbf{$dataset_T$} is divided into a training set and a test set, with the training set used to train the models and the test set used to evaluate their performance. Subsequently, we apply the trained models to \textbf{$dataset_A$} to assess their performance on previously unseen data, providing insights into the generalizability of the detection schemes. Next, we assess the analysis and explanation capabilities of ChatGPT (Section \ref{per_Chat}). Finally, we compare the performance of traditional models with that of ChatGPT (Section \ref{HE}).

\subsubsection{\textbf{Performance of Existing Methods}}\label{per_m}

\noindent The Android application is processed to extract both static and dynamic features, which are then provided as input to ChatGPT and used as training data for the model. \textbf{The Android application is not imported in its entirety for two principal reasons.} First, in order to maintain consistency in our experiments, it is necessary to align this solution with existing approaches that do not support the processing of the entire Android application as input. These methods typically concentrate on the analysis of particular features as opposed to the processing of complete binaries. This guarantees that the results obtained remain comparable across methods. Secondly, although ChatGPT is technically capable of reading binary files, such as those found in Android applications, it lacks the inherent ability to effectively interpret their complex structure and semantics. Binary files contain low-level machine code and compressed data that require specialized software to properly analyze. Direct input into a language model such as ChatGPT would not yield meaningful insights, as the model is optimized for processing human-readable text rather than machine-level instructions. Accordingly, we elected to extract pertinent features that ChatGPT is better equipped to process with greater efficiency, thereby enhancing the reliability and interpretability of the ensuing analysis.

In our experiments, we choose three state-of-the-art solutions: $Drebin$ \cite{r19}, $XM_{AL}$ \cite{wu}, and $MaMaDroid$ \cite{r25}. We select models that use both traditional machine learning and deep learning techniques to identify malicious applications. Additionally, these models use feature sets that contain the same categories as the feature sets we extract. 

We design a series of experiments to measure the efficacy of these three solutions in accurately identifying malicious software samples, as well as to compare them with ChatGPT. Below, we provide detailed insights into these methods.
\begin{itemize}
\item $Drebin$ \cite{r19} is a lightweight method proposed to address the security threat posed by malicious applications on the Android platform. As the proliferation of such applications hampers conventional defense mechanisms, Android smartphones often lack adequate protection against emerging malware. To address this issue, $Drebin$ performs an extensive static analysis by gathering a wide range of application features. These features are then embedded in a unified vector space, allowing for the identification of typical patterns indicative of malware. 

\item MaMaDroid \cite{r25} is a static analysis-based system designed to detect malware targeting the Android platform. $MaMaDroid$ takes a behavioral approach to malware detection by focusing on the sequence of abstracted application programming interface (API) calls performed by an app. The system abstracts the API calls into their corresponding class, package, or family, and constructs a model using the sequences obtained from the call graph of an app, treating them as Markov chains. This approach enhances the resilience of the model to changes in the underlying APIs and results in a manageable feature set.

\item $XM_{AL}$ \cite{wu} is a novel and interpretable machine learning-based method for detecting malware with high precision, while also explaining the detection results. $XM_{AL}$ uses multilayer perceptrons and attention mechanisms to identify the key features most relevant to the detection results. Then, $XM_{AL}$ automatically generates natural language descriptions to explain the core malicious behaviors within the application.
\end{itemize}

\begin{table}[t]
\centering
\caption{Results of Drebin, $XM_{AL}$ and MaMaDroid on $dataset_T$.}
\label{datasetT}
\begin{tabular}{ccccccc}
\hline
Method     & FN & FP  & Accuracy & Precision & Recall & F-Score \\ \hline
Drebin      & 5  & 4    & 0.955        & 0.963         & 0.951      & 0.952       \\
$XM_{AL}$      & 2  & 3    & 0.983        & 0.985         & 0.986      & 0.984       \\
MaMaDroid   & 2  & 1    & 0.979        & 0.978         & 0.973      & 0.981       \\ \hline
\end{tabular}

\end{table}

\begin{table}[t]
\centering
\caption{Results of Drebin, $XM_{AL}$ and MaMaDroid on $dataset_A$.}
\label{datasetA}
\begin{tabular}{ccccccc}
\hline
Method     & FN & FP  & Accuracy & Precision & Recall & F-Score \\ \hline
Drebin      & 9  & 3    & 0.914        & 0.903         & 0.911      & 0.912       \\
$XM_{AL}$     & 9  & 4    & 0.965        & 0.959         & 0.949      & 0.954       \\
MaMaDroid   & 6  & 11    & 0.945        & 0.939         & 0.940      & 0.941       \\ \hline
\end{tabular}

\end{table}

We use \textbf{$dataset_T$} to train these three schemes and test on \textbf{$dataset_A$}. The results are shown in Table \ref{datasetT} and Table \ref{datasetA}. 

As shown in Table \ref{datasetT}, $Drebin$, $XM_{AL}$, and $MaMaDroid$ all demonstrate impressive performance in \textbf{$dataset_T$}, with accuracy and F-Score exceeding 95\%. Meanwhile, the error rates of these three solutions are relatively low. Their FN values are 3, 2, and 2, respectively. These results prove the effectiveness of these Android malware detection solutions in accurately identifying malicious applications.

However, when we switch to a new dataset, the performance of $Drebin$, $XM_{AL}$, and $MaMaDroid$ decline, as shown in Table \ref{datasetA}. While $Drebin$, $XM_{AL}$, and $MaMaDroid$ maintain over 90\% in all four metrics on $dataset_A$, their FN values significantly increase. The FN values for $Drebin$, $XM_{AL}$, and $MaMaDroid$ increase from 3, 2, 2 to 9, 9, and 6 respectively. This indicates that these Android malware detection solutions suffer from dataset bias. Moreover, this suggests that these detection solutions have not truly understood the features we input; instead, they seem to primarily learn the unique statistical patterns inherent in datasets \cite{10232920}.

\begin{tcolorbox}[
  colframe = gray!30!white, colback = gray!10!white,
  colbacktitle = gray!30!white,
  coltext = black!50!black,
  coltitle = black!90!white]
{\setlength{\parindent}{0cm}
\textbf{Finding 1:} Existing detection solutions can effectively detect Android malware, but they suffer from dataset bias. Under a new dataset, the detection efficacy decreases, and the FN values significantly increase.}
\end{tcolorbox}

\subsubsection{\textbf{Performance of ChatGPT}}\label{per_Chat}
We use ChatGPT to analyze all samples in $dataset_A$, utilizing our designed prompts to guide ChatGPT. Meanwhile, the prompt we used is introduced in Section \ref{prompt} and illustrated in Figure \ref{fig:gpt1}.

\begin{figure}[htb]
\centering
\includegraphics[width=0.8\textwidth]{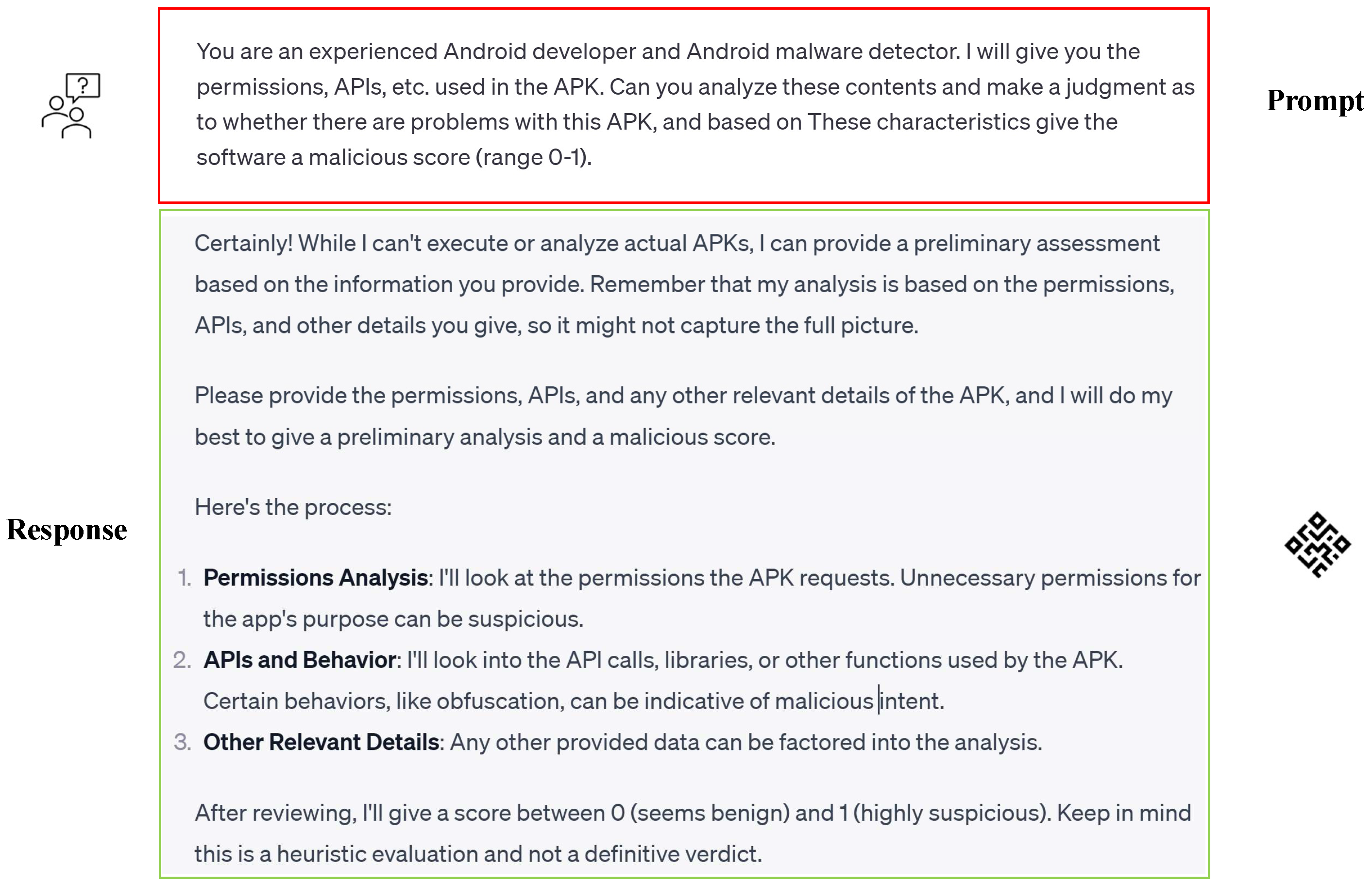}
\caption{\label{fig:gpt1}The initial guidance and outcome presentation from ChatGPT.}
\end{figure}

We use the designed prompts to guide ChatGPT, defining the tasks we need it to perform and the desired output. Meanwhile, due to OpenAI's limitations, ChatGPT cannot make definitive decisions. Therefore, we cannot use ChatGPT to determine whether a given sample is malicious or benign. Instead, ChatGPT can provide a maliciousness score for each sample, ranging from 0 to 1, where higher values indicate a higher degree of maliciousness.

We input various features of a sample into ChatGPT, enabling it to analyze the sample and generate results as shown in Figures \ref{fig:feature}, \ref{fig:sys}, \ref{fig:per}, and \ref{fig:score}.

\begin{figure}[htb]
\centering
\includegraphics[width=0.8\textwidth]{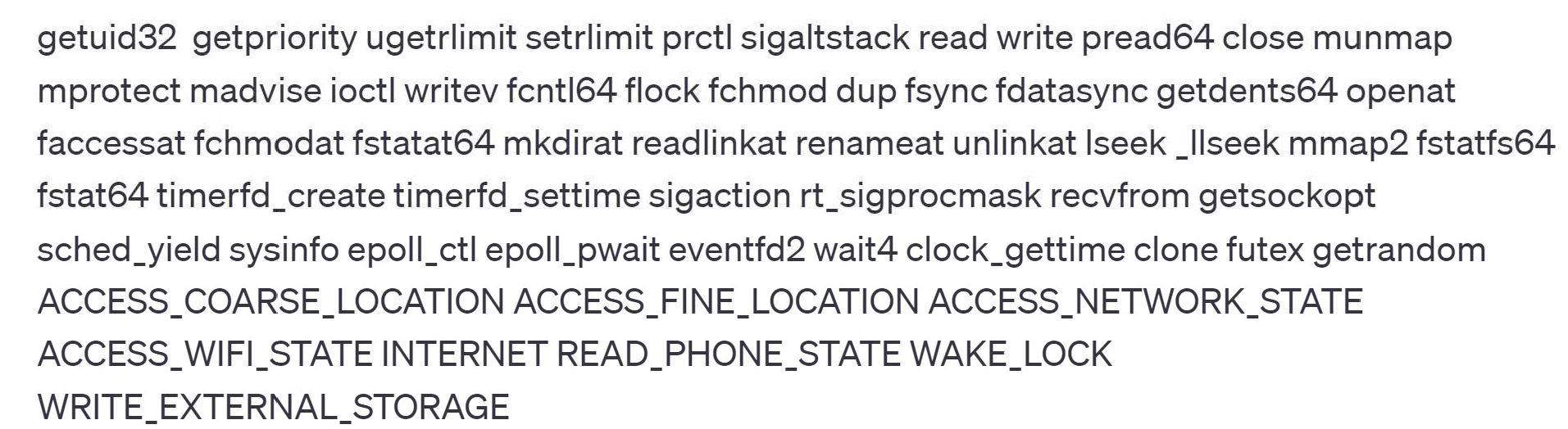}
\caption{\label{fig:feature}Input features, with each feature separated by a space.}

\end{figure}

We randomly select a malicious sample and input the static and dynamic features extracted from this sample into ChatGPT, as illustrated in Figure \ref{fig:feature}. Then ChatGPT provides analysis.

\begin{figure}[htb]
\centering
\includegraphics[width=0.8\textwidth]{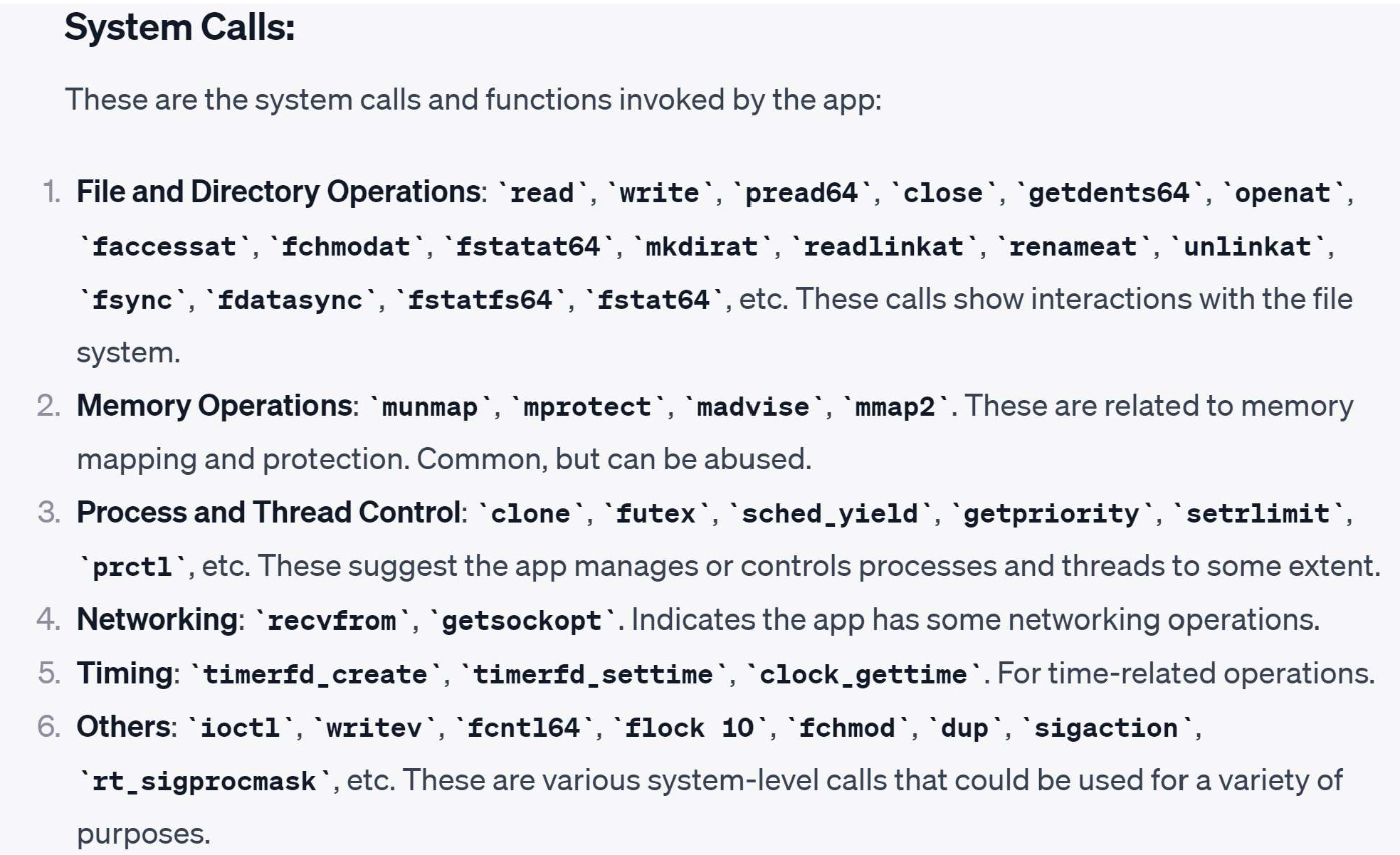}
\caption{\label{fig:sys}Analysis of system calls used by application.}

\end{figure}

First, ChatGPT analyzes the dynamic features employed by the sample, specifically system calls, as depicted in Figure \ref{fig:sys}. ChatGPT classifies these system calls into six categories and provides analysis for each category. These analyses indicate that the application uses the file system, abuses memory mapping, and performs networking operations. These insights inform developers in detail about the potential functionality of the application.

\begin{figure}[b]
\centering
\includegraphics[width=0.8\textwidth]{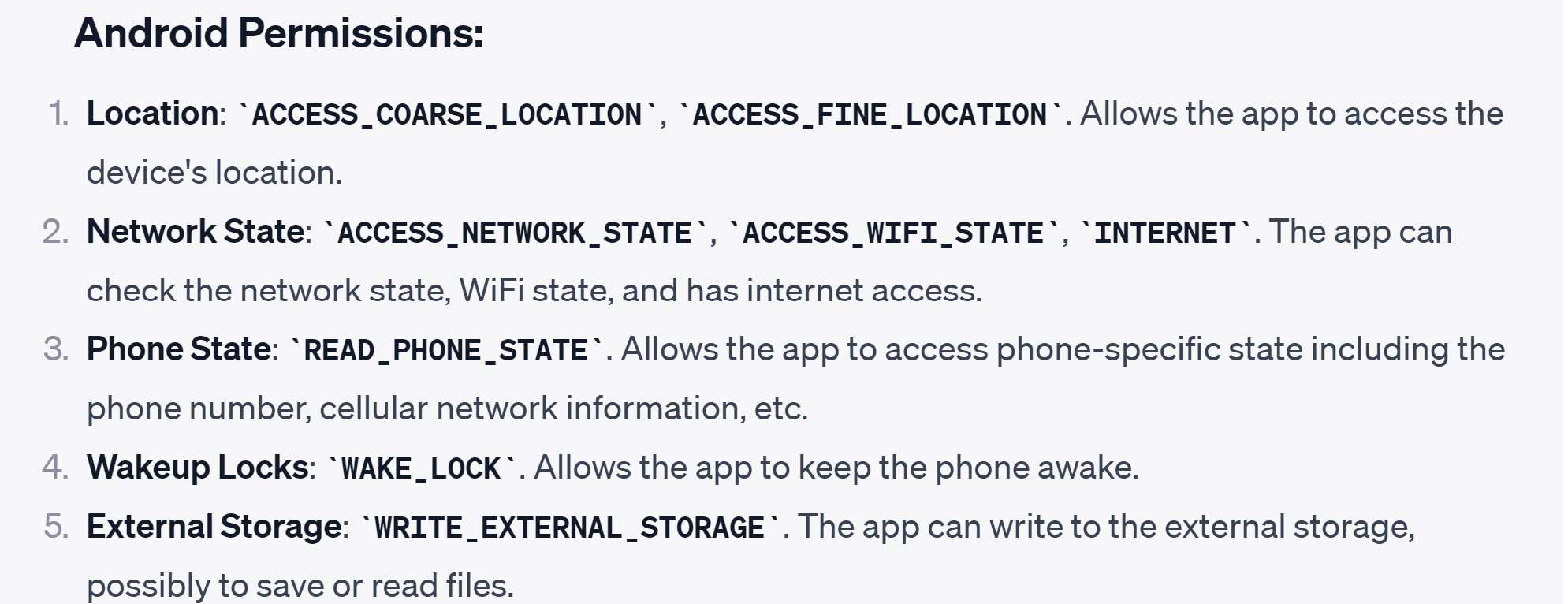}
\caption{\label{fig:per}Analysis of permissions used by application.}
\end{figure}

Second, ChatGPT analyzes the permissions declared by the application. ChatGPT categorizes these permissions into five distinct categories and provides explanations for each category, as depicted in Figure \ref{fig:per}. These results reveal that the application can access the device's location, retrieve network status, wake up the device, and access device identification, among other capabilities. This provides developers with a clear understanding of the permissions sought by the application.

\begin{figure}[htb]
\centering
\includegraphics[width=0.8\textwidth]{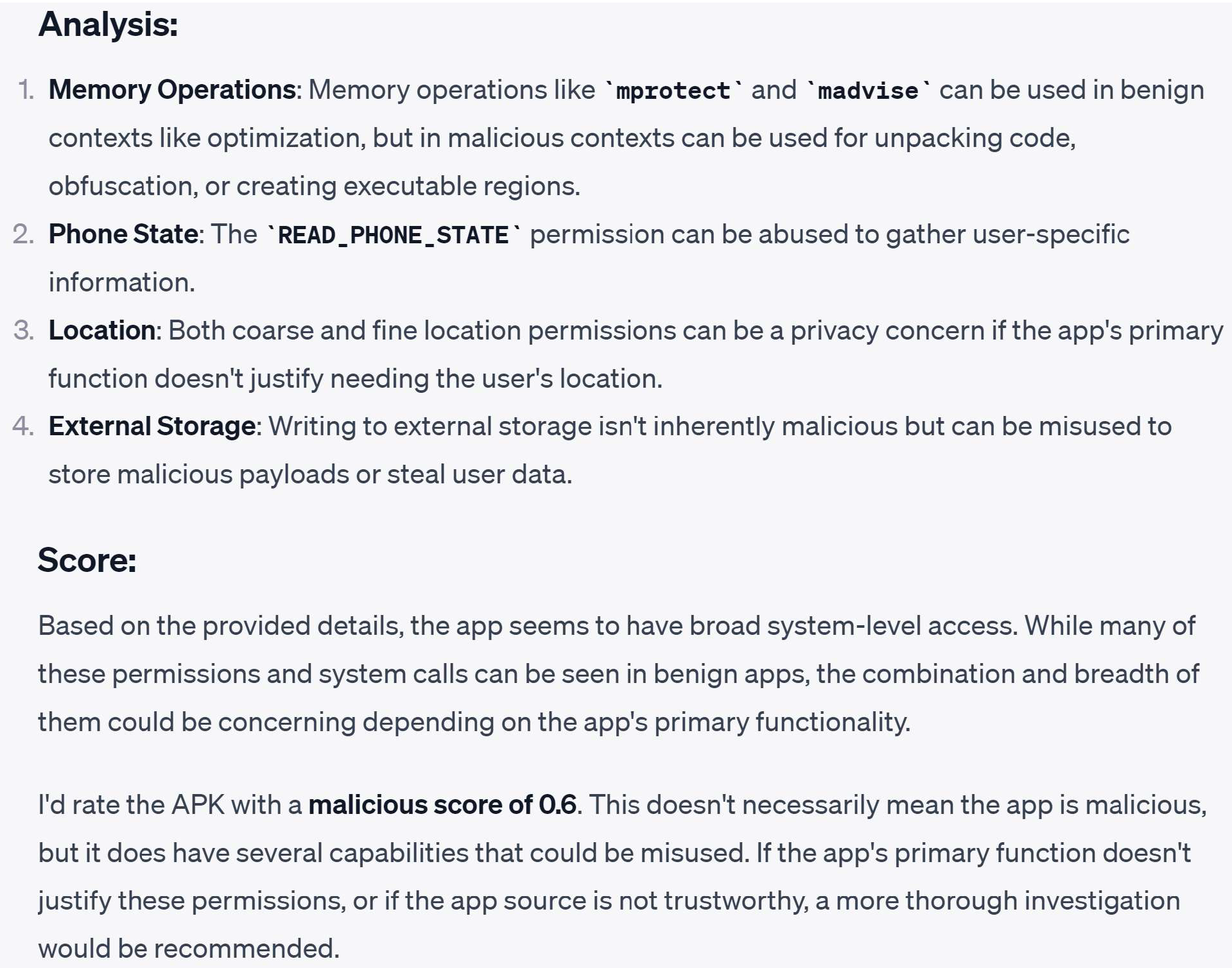}
\caption{\label{fig:score}Summary analysis and the assignment of maliciousness scores.}

\end{figure}

Finally, ChatGPT summarizes the aforementioned feature analysis, underscores four potentially malicious behaviors, and assigns a degree of maliciousness to the application.

As illustrated in Figures \ref{fig:sys}, \ref{fig:per}, and \ref{fig:score}, ChatGPT thoroughly examines the features within each category, providing explanations regarding their functions and potential security threats. It is important to note that, owing to the constraints of OpenAI, ChatGPT cannot provide definitive decisions but can only offer an estimate of the potential level of risk \cite{gpt4}. In contrast to existing detection solutions, ChatGPT does not provide definitive decisions, which means we cannot directly determine whether a piece of software is malware. However, ChatGPT offers an extensive analysis, furnishing explanations for each feature within its respective category and delving into potential threats associated with these features. These detailed analyses and explanations help enhance developers' understanding. They enable developers to gain a comprehensive grasp of the functionality utilized by the software and its potential implications. 

Moreover, we compare the detailed reports provided by ChatGPT with the explanatory descriptions of $XM_{AL}$ from $Drebin$. $Drebin$ simply provides basic definitions for each feature. $XM_{AL}$ deduces features that make decisions based on detection results and provide natural language descriptions. Both $Drebin$ and $XM_{AL}$'s explanatory descriptions lack analysis and more detailed, in-depth descriptions. Conversely, while ChatGPT may not offer specific decisions, it conducts analysis and provides explanations based on features, deducing results from underlying causes rather than reverse-engineering causes from outcomes. This illustrates ChatGPT's genuine understanding of input features and its analytical approach.

\begin{tcolorbox}[
  colframe = gray!30!white, colback = gray!10!white,
  colbacktitle = gray!30!white,
  coltext = black!50!black,
  coltitle = black!90!white]
{\setlength{\parindent}{0cm}
\textbf{Finding 2:} Existing detection solutions either provide only decisions or generate reports that are mere explanations of features. In contrast, ChatGPT not only offers detailed explanations but also provides in-depth analysis.}
\end{tcolorbox}

\subsubsection{\textbf{Human Evaluation}}\label{HE}
In order to conduct a comprehensive comparison and analysis between ChatGPT and existing detection solutions, we invite ten experienced developers, each with more than 10 years of experience in Android, to participate in a questionnaire. All ten respondents are engaged in Android mobile software development, which aligns closely with the focus of our study. Their work involves various aspects of the Android ecosystem, including coding, testing, and maintaining mobile applications. This hands-on experience ensures they possess in-depth knowledge of Android development frameworks, tools, and best practices. The specific questions are as follows:
\begin{enumerate}
    \item \textbf{Android Device Usage:}
    \begin{itemize}
        \item Do you frequently utilize devices powered by the Android operating system, such as smartphones and tablets?
    \end{itemize}
    \item \textbf{Interactions with Antivirus Software:}
    \begin{itemize}
        \item How often do you come across alerts or detections from antivirus software when using or installing applications on your Android device? 
        \begin{itemize}
            \item Rate your experience on a scale of 0 to 5:
            \item 0 - Never encountered this situation;
            \item 5 - Frequently encounter this situation.
        \end{itemize}
    \end{itemize}
    \item \textbf{Evaluation of Existing Detection Solutions:}
    \begin{itemize}
        \item We provide three state-of-the-art Android malware detection solutions, each with 20 detection reports. Please select five reports from each of these three solutions. After reading these reports, please answer the following questions: Do you think the reports provided by the state-of-the-art Android malware detection solutions are detailed? Are you concerned about the reliability of the decision-making process of these solutions?
    \end{itemize}
    \item \textbf{Understanding ChatGPT’s Analysis:}
    \begin{itemize}
        \item We provide 20 detection reports generated by ChatGPT. Please randomly select five reports from them. After reading them, please answer the following question: Do you think the reports provided by the ChatGPT are detailed? Are concerned that ChatGPT cannot provide specific decisions?
    \end{itemize} 
    \item \textbf{Comparative Report Preferences:}
    \begin{itemize}
        \item After reviewing both sets of reports, which format do you find more appealing and why? To assist in our analysis, kindly rate the two types of reports based on: 
        \begin{itemize}
            \item Comfort: How easily do you feel when trying to understand the content?
            \item Readability: How clear and understandable is the text?
            \item Friendliness: How approachable and user-friendly is the content?
        \end{itemize}
        \item For each aspect, use a scale of 0 to 5, where:
        \begin{itemize}
            \item 0 signifies extremely poor;
            \item 5 denotes exceptional performance.
        \end{itemize}
    \end{itemize}
\end{enumerate}

We consolidate feedback from all participants, which is illustrated in Figures \ref{fig:q1score}, \ref{fig:q5amd}, and \ref{fig:q5gpt}.

\begin{figure}[htb]
\centering
\includegraphics[width=0.8\textwidth]{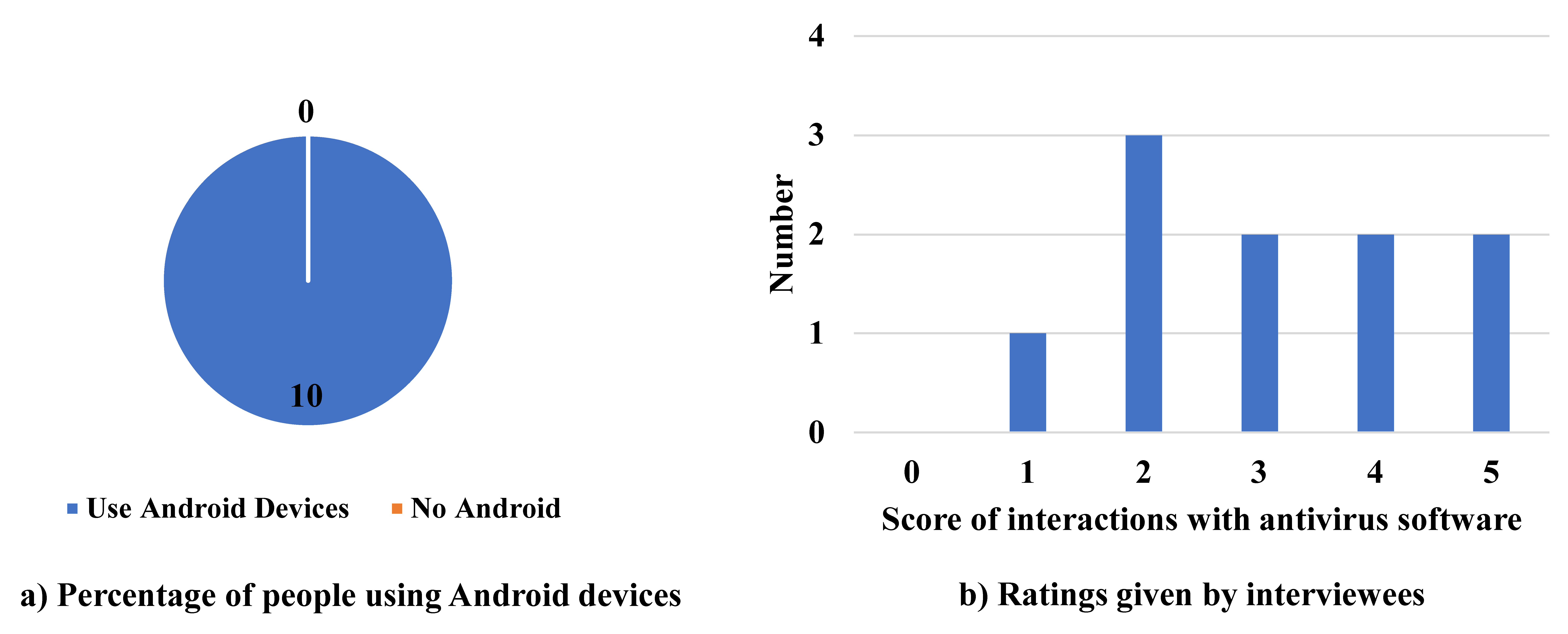}
\caption{\label{fig:q1score}Percentage of participants who use Android devices and the regularity of their interactions with antivirus software.}

\end{figure}
For the first question, ten experienced developers state that they often use Android-based devices, such as mobile phones and tablets. For the second question, to facilitate quantitative analysis, we ask respondents to rate it. As shown in Figure \ref{fig:q1score}, four respondents scored below 3, while the remaining 6 respondents scored 3 or above. This suggests that, both in professional settings and in daily life, people frequently use antivirus software on Android devices and frequently encounter diagnostic reports generated by these software.

For the third question, the ten developers have varying opinions on the three detection solutions. First, all ten developers unanimously find the reports generated by $MaMaDroid$ to be very concise, providing only detection results. They also express significant concerns about how $MaMaDroid$ makes its decisions. Second, 30\% of the respondents believe that $Drebin$'s reports are relatively detailed but of limited utility. The remaining 70\% of the respondents find $Drebin$'s reports not very detailed, merely providing explanations of some feature definitions with no practical value. Additionally, all ten respondents agree that $Drebin$'s reports do not reveal how it arrives at its judgments, and they still have concerns about the reliability of $Drebin$'s decision-making process. Third, the ten respondents believe that compared to $MaMaDroid$ and $Drebin$, $XM_{AL}$'s reports contain more content and have clearer descriptions. However, they feel that $XM_{AL}$ still does not provide analysis and explanations; it merely describes the functionality of the features being used. Based on $XM_{AL}$'s reports, they also cannot ascertain the basis for its decision-making process, leading to concerns about its reliability.

\begin{figure}[h]
\centering
\includegraphics[width=1.0\textwidth]{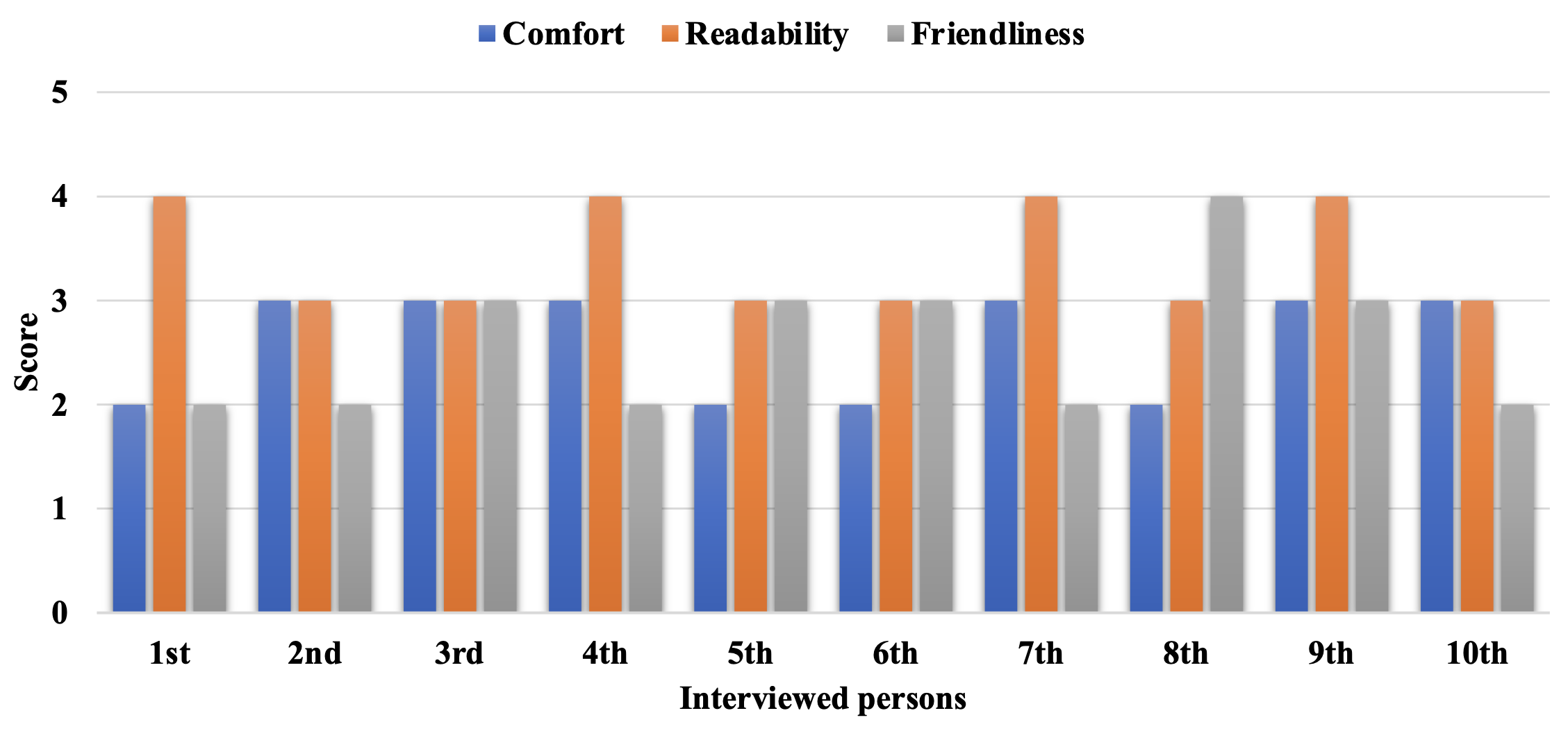}
\caption{\label{fig:q5amd} Respondents' feedback on reports of existing detection solutions. The vertical axis displays the scores, ranging from 0 to 5, while the horizontal axis represents the respondents, numbered sequentially from the first to the tenth. }
\end{figure}

\begin{figure}[h]
\centering
\includegraphics[width=1.0\textwidth]{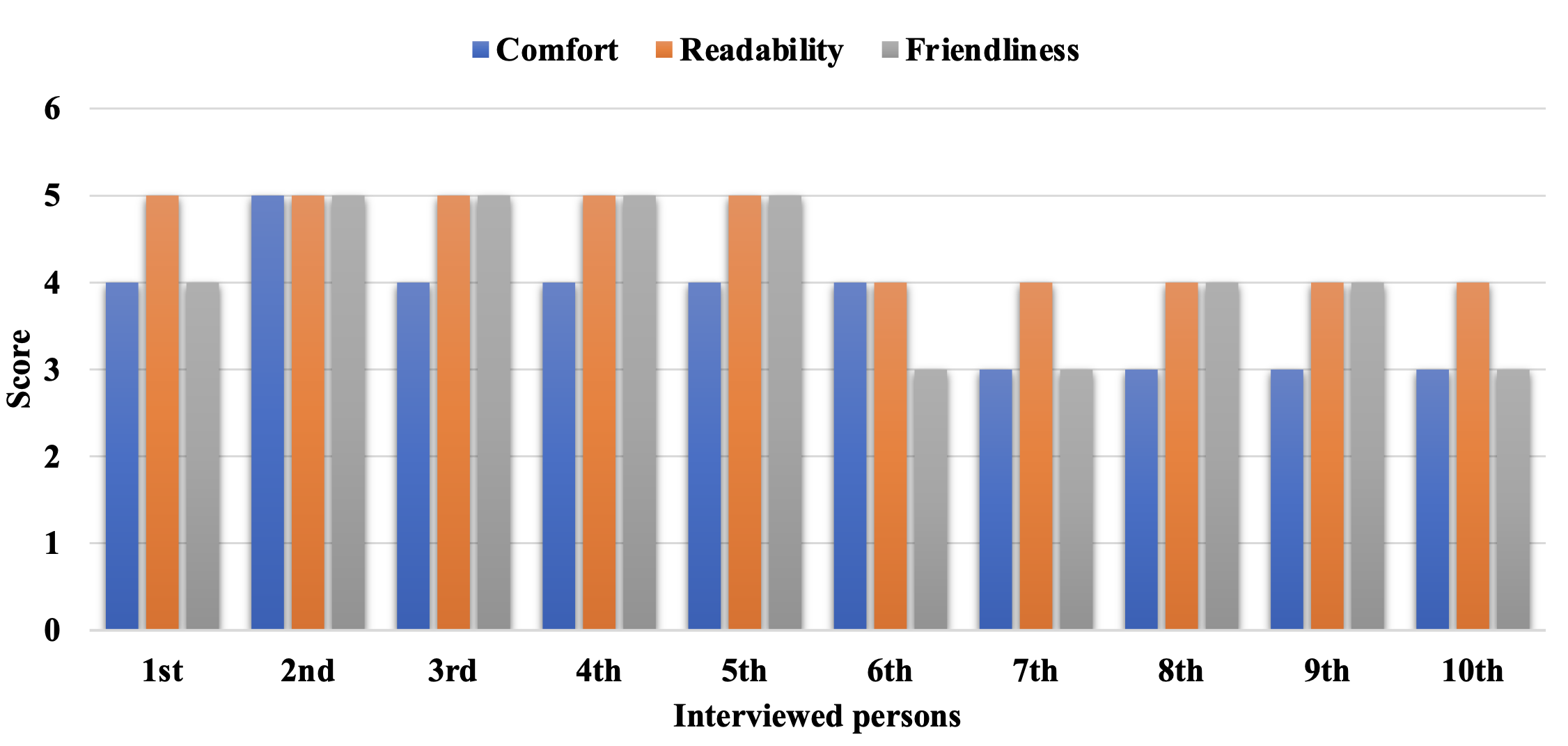}
\caption{\label{fig:q5gpt}Respondents' feedback on reports of ChatGPT. The vertical axis represents the score, while the horizontal axis represents the respondents. The vertical axis displays the scores, ranging from 0 to 5, while the horizontal axis represents the respondents, numbered sequentially from the first to the tenth.}
\end{figure}

For the fourth question, we summarize the feedback from the ten respondents. First, ten developers agree that the reports provided by ChatGPT are detailed, including not only analysis and explanations of input features but also their potential implications and final summary analysis. However, ten developers also state that ChatGPT's inability to provide decisions and judgments is a major drawback. They mention that if ChatGPT needs further improvement, it should be able to make decisions.

In reference to the fifth question, every interviewee expresses a unanimous preference for ChatGPT due to its provision of clear explanations. Furthermore, we ask the interviewees to assess the decisions made by existing models and the explanations provided by ChatGPT, evaluating them on three key aspects—\textit{Comfort, Readability, and Friendliness}, as illustrated in Figures \ref{fig:q5amd} and \ref{fig:q5gpt}. From these two figures, it is evident that all ten interviewees have given higher ratings to the analysis and explanations offered by ChatGPT. This reinforces the notion that ChatGPT's explanations are more intuitive and easier to comprehend.

The survey responses offer insightful perspectives on Android malware detection solutions and the preferences of experienced developers. Most respondents, who frequently use Android-based devices both professionally and personally, demonstrate a high reliance on antivirus software. This underscores the importance of effective malware detection tools in real-world applications.

In contrast, ChatGPT's reports were universally praised for their clarity, detail, and comprehensive analysis, making them more intuitive and user-friendly. However, respondents also noted its limitation in providing final decisions or judgments, which is a significant drawback for practical application in malware detection. Despite this, the developers overwhelmingly preferred ChatGPT for its readability, comfort, and overall user-friendliness, reinforcing the importance of clear and understandable explanations in building user trust. Overall, while existing Android malware detection models are widely used, their lack of transparency and interpretability remains a key challenge. ChatGPT’s strength lies in its ability to offer detailed explanations, making it a valuable tool for developers, though further development is needed to enable it to provide definitive decisions.

\begin{tcolorbox}[
  colframe = gray!30!white, colback = gray!10!white,
  colbacktitle = gray!30!white,
  coltext = black!50!black,
  coltitle = black!90!white]
{\setlength{\parindent}{0cm}
\textbf{Finding 3:} Compared to existing detection solutions, the reports provided by ChatGPT are not only detailed but also highly readable. However, ChatGPT has a drawback in that it cannot make decisions.}
\end{tcolorbox}

\subsection{Influence and improvement}\label{impact}

\begin{tcolorbox}[title = {RQ 2:},
  colframe = gray!30!white, colback = gray!10!white,
  colbacktitle = gray!30!white,
  coltext = black!50!black,
  coltitle = black!90!white]
  How does the large language model (i.e., ChatGPT) impact existing Android malware detection solutions? And what valuable insights can it provide to motivate developers?
\end{tcolorbox}

\noindent In our previous discussions, we demonstrated the effectiveness of existing detection solutions in detecting malicious software through a series of experiments. However, we also discover that despite their strong performance and high detection rates, they are still susceptible to dataset biases. Additionally, these solutions suffer from low interpretability. The reports they provide are often simple explanations of features, and some solutions only provide detection results. Developers cannot discern the reasons and basis for the decisions made by these solutions from these reports and detection results, which has raised concerns about their decision-making reliability.

In contrast, compared to these decision-based detection solutions, ChatGPT, although unable to make explicit decisions, can provide a wealth of rich and detailed analysis and explanations. These analyses and explanations allow developers to gain deeper insights into the functionality used by the software and potential issues it may pose. Therefore, in order to improve and enhance detection solutions, we invite ten experienced developers to participate in a questionnaire. Six of the ten respondents are actively involved in antivirus software development, contributing their specialized expertise in identifying and mitigating security threats. Three respondents are engaged in artificial intelligence development, including predictive analytics, machine learning, and intelligent automation. The remaining respondent focuses on Android software development. Their expertise adds valuable context to the analysis, particularly in examining the practical implications of malware detection systems within the Android environment. We prepare four questions to gain a deeper understanding:
\begin{enumerate}
    \item Are you acquainted with the current leading Android malware detection tools? How would you rate their beginner-friendliness on a scale of 0 to 5? 
    \item From your experience, which facets of the existing Android malware detection solutions could be enhanced? Please provide ratings on a scale for the following dimensions:
    \begin{itemize}
        \item Interpretability: How clear and comprehensible are the results or outputs?
        \item Ease of Use: How user-friendly and intuitive is the software or tool?
        \item Convenience: How streamlined and efficient is the detection process?
        \item For each aspect, use a scale of 0 to 5, where:
        \begin{itemize}
            \item 0 denotes no improvement is needed;
            \item 5 means improvement is urgently needed.
        \end{itemize}
    \end{itemize}
    
    \item What are the advantages and disadvantages of non-decision-based large language models like ChatGPT compared to traditional Android malware detection tools that rely on decision-making?
    \item Would you be open to adopting a specialized large language model designed for Android malware detection if it becomes available in the foreseeable future?
\end{enumerate}

For the first question, everyone has exposure to or has used the current advanced Android malware detection solutions. We also ask them to rate the ease of using these solutions. 20\% of the people give them 4 points, 50\% of the people give them 3 points, and 30\% of the people give them 2 points. This indicates that only 20\% of respondents currently perceive existing Android malware detection solutions to be user-friendly, while the remaining 80\% encounter a range of difficulties. The reported difficulties include the presence of complex interfaces and a paucity of user-friendly documentation, which collectively render these tools inaccessible to non-experts. Furthermore, a considerable proportion of respondents indicated that the replication or deployment of these solutions necessitates a considerable investment of time and effort, frequently due to intricate setup procedures, dependency on particular environments, or the necessity for extensive technical expertise. This indicates a clear necessity for the development of more intuitive and streamlined tools, with the objective of reducing the barrier to entry for users and enhancing the overall efficiency of deployment and use.

\begin{tcolorbox}[
  colframe = gray!30!white, colback = gray!10!white,
  colbacktitle = gray!30!white,
  coltext = black!50!black,
  coltitle = black!90!white]
{\setlength{\parindent}{0cm}
\textbf{Finding 4:} The deployment and replication of existing Android malware detection solutions frequently presents significant challenges, primarily due to their intricate setup procedures and the necessity for a highly specialized technical infrastructure. These tools necessitate the navigation of numerous dependencies, the configuration of intricate parameters, and the resolution of compatibility issues between diverse devices or systems. Furthermore, the absence of transparent, user-friendly documentation exacerbates the complexity, rendering it challenging for users (particularly those lacking profound technical expertise) to effectively utilize these solutions.}
\end{tcolorbox}

For the second question, we ask respondents to evaluate whether existing solutions need improvement based on three aspects: interpretability, ease of use, and convenience. They were required to rate these three aspects on a scale from 0 to 5, where 0 represents no need for improvement and 5 represents an urgent need for improvement. Detailed results are shown in Figure \ref{fig:q2amd}. From Figure \ref{fig:q2amd}, we can observe that the ten developers are more concerned about the interpretability and convenience of the detection solutions. In terms of interpretability, two developers give a rating of five, indicating an urgent need for improvement. The rest of the developers give scores of no less than three. Regarding convenience, three developers give a rating of five, and the remaining developers also give scores of three or above. This suggests that developers prefer a detection tool that is highly interpretable and readily accessible.

\begin{figure}[htb]
\centering
\includegraphics[width=1.0\textwidth]{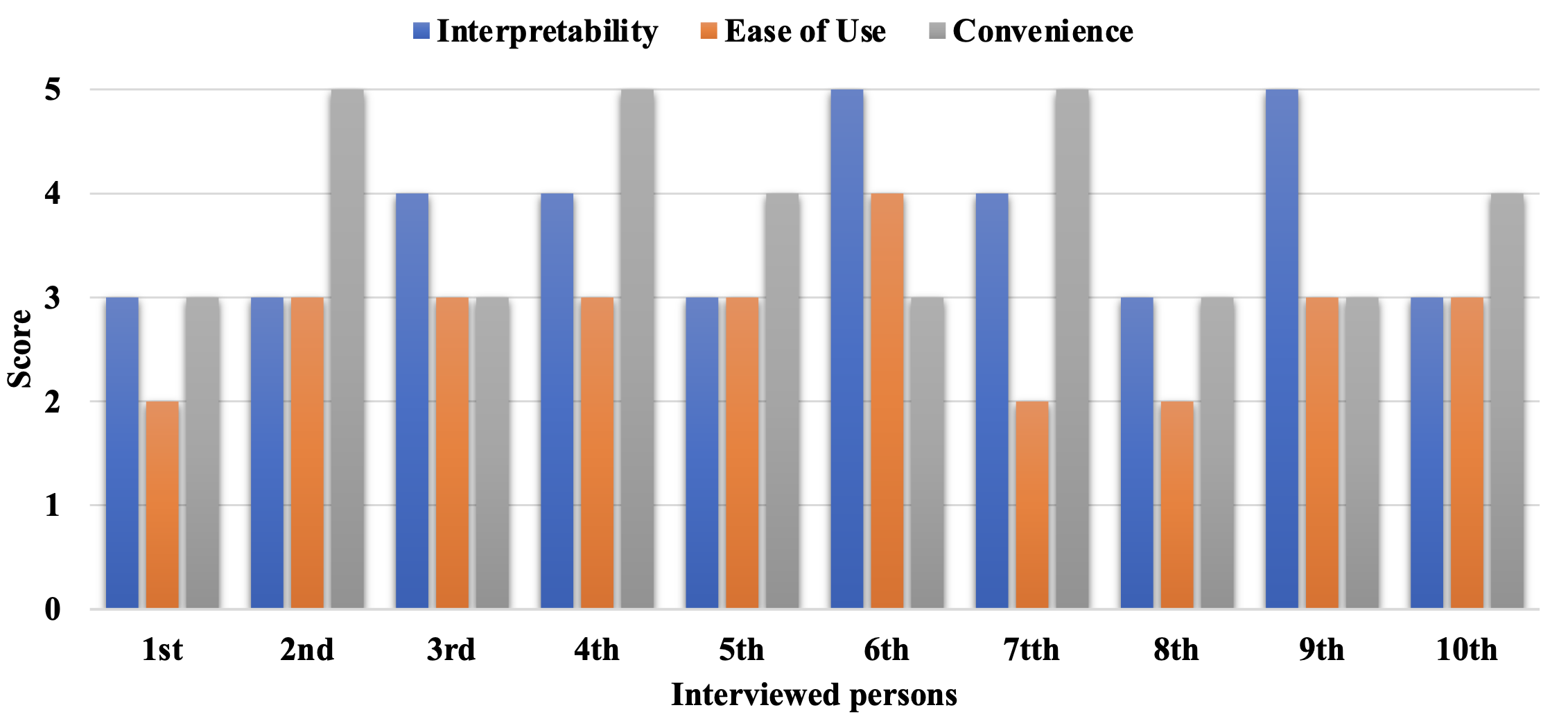}
\caption{\label{fig:q2amd} Respondents' feedback on potential enhancements of existing Android malware detection solutions. The vertical axis displays the scores, ranging from 0 to 5, while the horizontal axis represents the respondents, numbered sequentially from the first to the tenth.}
\end{figure}

\begin{tcolorbox}[
  colframe = gray!30!white, colback = gray!10!white,
  colbacktitle = gray!30!white,
  coltext = black!50!black,
  coltitle = black!90!white]
{\setlength{\parindent}{0cm}
\textbf{Finding 5:} For existing Android malware detection solutions, interpretability and convenience need to be further improved.}
\end{tcolorbox}

For the third question, we summarize the feedback from the ten developers. First, ten developers agree that, compared to existing detection solutions, ChatGPT's primary advantage is its robust analytical and explanatory capabilities. ChatGPT can comprehensively analyze and explain input features. Second, ChatGPT's ease of use is another advantage. The conversational input and output format make it more convenient for developers to use. Meanwhile, all ten respondents also mention what they consider to be the main drawback, which is ChatGPT's inability to make absolute decisions. This is seen as a critical flaw for Android malware detection tasks that rely heavily on decision-making. Meanwhile, regarding the fourth question, ten respondents express their anticipation for large language models specifically designed for Android malware detection.

The survey results reveal significant insights into the current state of Android malware detection solutions and highlight key areas for improvement. While all respondents have experience with detection tools, the majority find them challenging to use, with only 20\% rating them as user-friendly. The remaining 80\% encounter difficulties such as complex interfaces, a lack of user-friendly documentation, and time-consuming setup processes, which create barriers to accessibility, especially for non-experts. 

Regarding ChatGPT, all respondents acknowledged its strengths in analytical capabilities and ease of use, particularly its ability to provide clear, detailed explanations in a conversational format. However, the inability of ChatGPT to make definitive decisions was identified as a major drawback, as decision-making is crucial in malware detection tasks. 

In conclusion, the survey highlights the need for more user-friendly and interpretable malware detection tools that not only simplify the user experience but also provide clear insights into decision-making processes. ChatGPT shows promise due to its analytical strengths and ease of use, but its inability to make concrete decisions indicates the need for further development to make it a more practical tool for malware detection. The future of Android malware detection lies in integrating robust explanatory features with actionable decision-making capabilities, ensuring both accessibility and reliability for developers.

\begin{tcolorbox}[
  colframe = gray!30!white, colback = gray!10!white,
  colbacktitle = gray!30!white,
  coltext = black!50!black,
  coltitle = black!90!white]
{\setlength{\parindent}{0cm}
\textbf{Finding 6:} While ChatGPT possesses strong analytical and explanatory capabilities, it falls short in delivering conclusive decisions.}
\end{tcolorbox}

\subsection{Future development}\label{improve}

\begin{tcolorbox}[title = {RQ 3:},
  colframe = gray!30!white, colback = gray!10!white,
  colbacktitle = gray!30!white,
  coltext = black!50!black,
  coltitle = black!90!white]
  How to enhance the Android Malware Detection capability of existing large language models (i.e., ChatGPT)?
\end{tcolorbox}

\noindent After the previous research and investigation, we find that although ChatGPT can provide detailed analysis and explanation, it is incapable of making the final decision, that is, determining whether an application is benign or malicious. In fact, we construct the prompt, trying to get ChatGPT to make a definitive decision or judgment, but it refuses each time. ChatGPT even resists giving a rating on the level of maliciousness. Through continuous optimization and testing, we determine the appropriate prompt that can elicit a maliciousness level rating from ChatGPT. This is precisely a limitation of ChatGPT. ChatGPT, constrained by OpenAI's requirements, is unable to make definitive judgments \cite{gpt4}. But We want large models like ChatGPT to be capable of both making decisions and providing analysis and explanations. \textbf{Based on this, we hope to provide developers with more insights from a non-decision perspective to facilitate future improvements and the development of better Android malware solutions.} Therefore, our improvement plan is divided into two steps: 1) Improving ChatGPT; 2) Constructing a large language model specifically for Android malware detection.

First, to enhance ChatGPT, we can provide more data to allow it to analyze more deeply. For this, we can ask ChatGPT what additional data it needs to further enhance the analysis and interpretation of Android malware. Therefore, we can inquire with ChatGPT about what additional data it requires to further strengthen its analysis and interpretation of malicious Android applications. After inquiring, we learn that providing the hardware resources used by the APK, Intent, Activity, shell commands, and background information of the application can allow ChatGPT to better analyze Android malicious applications. Meanwhile, we can build corresponding retrieval enhancement generation (RAG) for ChatGPT. We use the open source RAG framework\footnote{TinyRAG, \url{https://github.com/phbst/tinyRAG/tree/main}}. Then we collect Android permissions\footnote{Adnroid permission, \url{https://developer.android.com/guide/topics/permissions/overview?hl=en}} and APIs\footnote{Android API, \url{https://developer.android.google.cn/reference/packages}} to form a vector database. 

\begin{figure}[htb]
\centering
\includegraphics[width=0.8\textwidth]{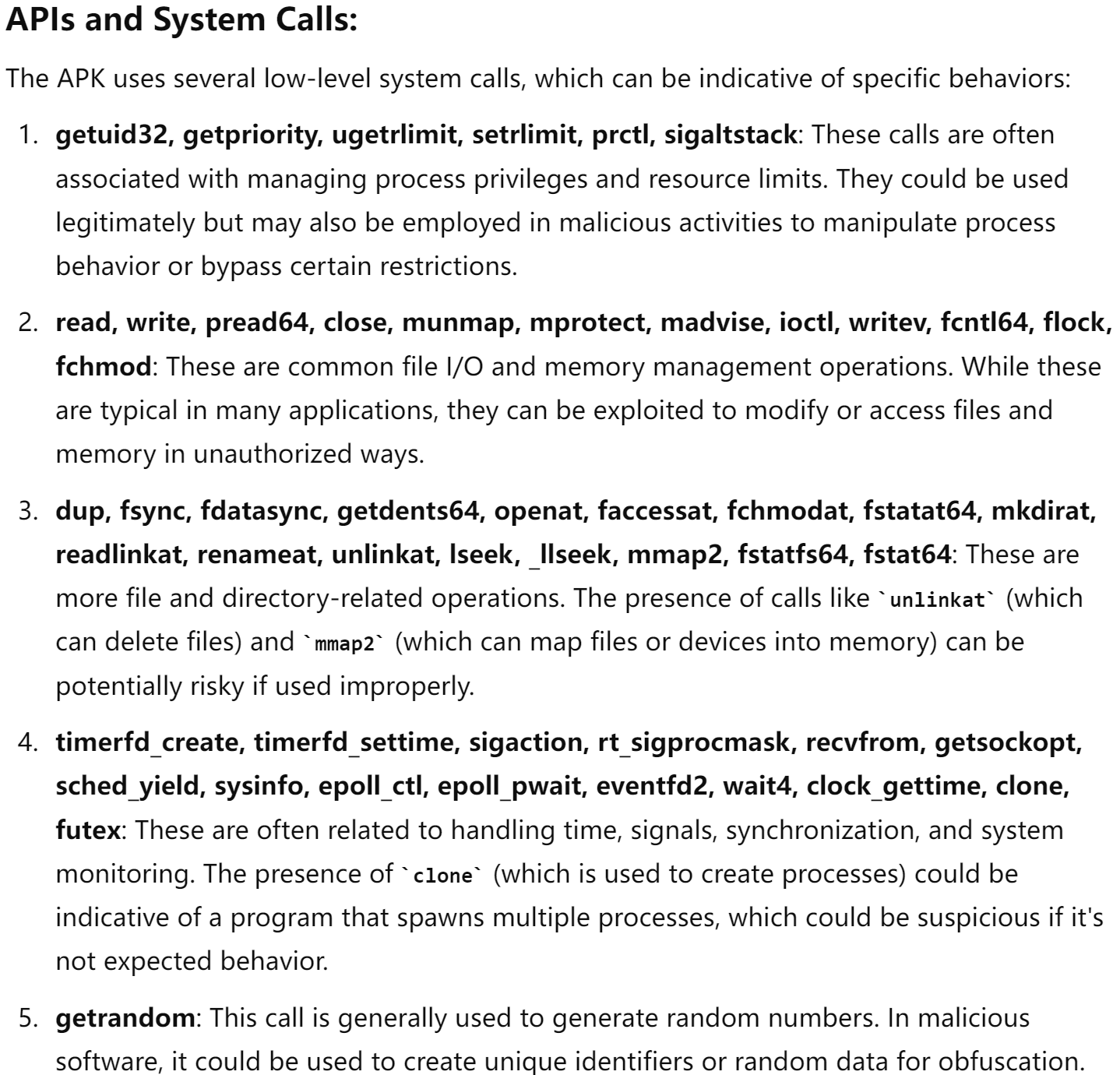}
\caption{\label{fig:rag1} Analysis of API and system calls used by the application.}
\end{figure}

\begin{figure}[htb]
\centering
\includegraphics[width=0.8\textwidth]{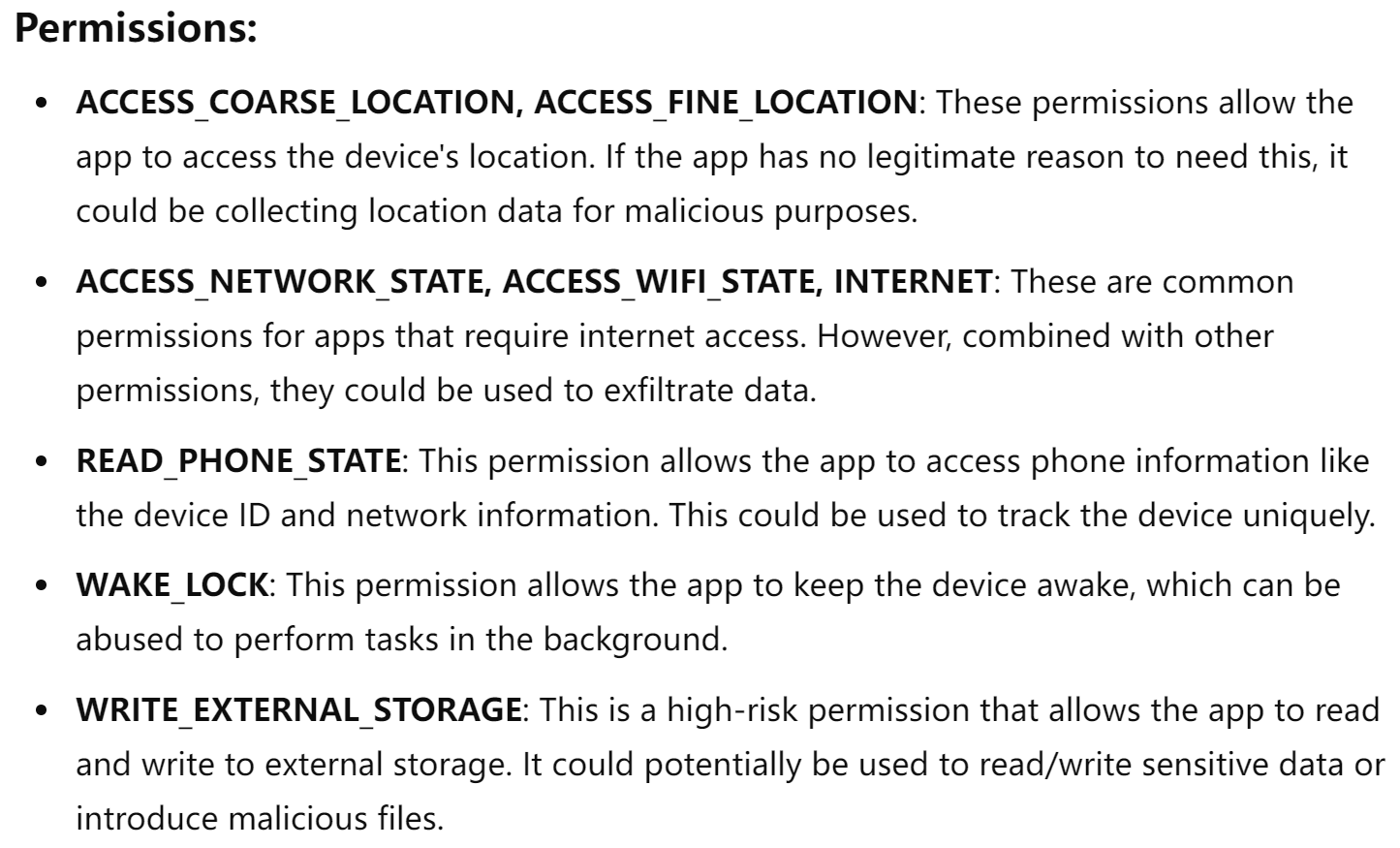}
\caption{\label{fig:rag2} Analysis of permission used by the application.}
\end{figure}

\begin{figure}[htb]
\centering
\includegraphics[width=0.8\textwidth]{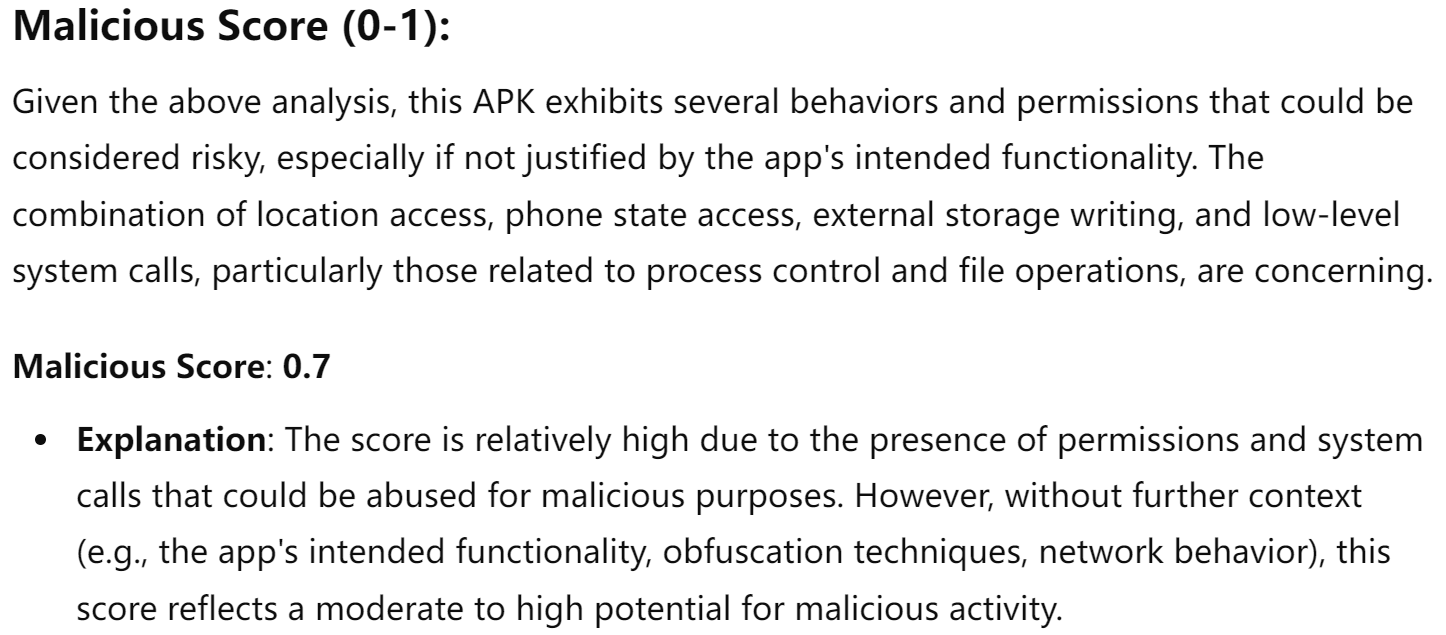}
\caption{\label{fig:rag3} Summary analysis and the assignment of maliciousness scores.}
\end{figure}

Figures \ref{fig:rag1}, \ref{fig:rag2}, and \ref{fig:rag3} show ChatGPT’s responses after adding RAG. The example shown here is the same as that used in Section \ref{relia}. At the same time, the same prompt is still used for booting. Then, we interview fifty participants to assess whether RAG enhances their experience with ChatGPT. The opinions of the fifty participants can be categorized into three groups. Thirty-six participants feel that while RAG offered some improvement, it is not substantial. Eleven participants believe that RAG provided more detailed explanations. The remaining three participants think there is no improvement at all. Limited by the design of the retrieval algorithm and the construction of the database, although the simple RAG has a certain improvement over ChatGPT, it does not have a better performance. 

Although the analysis using RAG yields a greater number of results compared to the analysis without RAG, each feature is described in significantly more detail. However, the score for the same malicious application shows only a slight increase, from 0.6 to 0.7. This suggests that while ChatGPT's understanding of the characteristics leads to a more thorough analysis, the overall assessment score remains relatively stable. This demonstrates that ChatGPT, when enhanced with RAG, can provide more comprehensive insights into each feature, but the impact on the final scoring is modest.

Second, to construct a large language model focused on Android malware detection, we can use open-source large models for retraining. Through the first step, we will collect a large amount of feedback from ChatGPT, which can be used for model fine-tuning. Moreover, we need to collect a large number of benign and malicious samples, along with background and contextual information about these samples.

\begin{tcolorbox}[
  colframe = gray!30!white, colback = gray!10!white,
  colbacktitle = gray!30!white,
  coltext = black!50!black,
  coltitle = black!90!white]
{\setlength{\parindent}{0cm}
\textbf{Finding 7:} The enhancement goal is for large models to provide analysis and explanations while making decisions. Therefore, the improvement plan is divided into two steps: first, enhancing ChatGPT, and second, constructing a large model for Android malware detection.}
\end{tcolorbox}
\section{Discussion}\label{discuss}
\subsection{Analysis}
\noindent Our study of the impact of ChatGPT and LLM on Android malware detection challenges traditional decision-based detection models and proposes a shift towards more easily interpretable, analysis-driven approaches. \textbf{This shift is significant because traditional models, while effective, often lack transparency in the decision-making process, which can impede developers' ability to trust and enhance the detection process.}

The potential for enhancing the interpretability and analysis of malware detection is expanding with the advent of LLMs, such as ChatGPT. LLMs can furnish comprehensive explanations, furnish insight into the rationale behind decision-making processes, and assist in identifying patterns of malware behavior that may not be readily discernible through conventional methods. This enables not only the detection of malware but also the understanding of the reasoning behind each detection decision. This, in turn, paves the way for hybrid models where decisions are supported by deeper analysis and explanation.

LLMs have the potential to transform malware detection by increasing the visibility of the models' detection processes, thereby bridging the gap between decision-making and interpretability. \textbf{This transition from a decision-oriented to an explanation-oriented framework has the potential to markedly enhance confidence in detection models, facilitate more efficacious enhancements, and culminate in the development of more sophisticated malware detection systems.}

\textbf{Our research indicates that LLM (e.g., ChatGPT) has made notable progress in the interpretability and accuracy of recognition systems.} While traditional decision-based models are undoubtedly powerful, they are often constrained by a lack of transparency and an inability to elucidate the underlying decision-making processes. This study demonstrates the advantages of integrating LLMs to enhance not only the detection process but also the transparency and credibility of these systems.

\textbf{A principal finding of this research is the necessity for interpretability in the context of malware detection.} The operation of traditional models as "black boxes" makes it challenging for developers and analysts to comprehend the underlying decision-making process. The integration of LLMs enables a more transparent and interpretable approach to malware detection, facilitating a deeper understanding of the underlying decision-making process and the rationale behind the identification of specific threats among users.

Further research should concentrate on enhancing the integration of LLM with conventional malware detection techniques. One avenue of promising research is the development of hybrid models. Furthermore, it is imperative to examine the efficacy of these models in actual operational contexts, particularly in regard to their scalability, computational demands, and practical deployment.

Furthermore, future research should examine how the more sophisticated analytical abilities of the LLM can be leveraged to minimize false positives and omissions. This may entail fine-tuning the model to gain a deeper understanding of the contextual factors that contribute to malware behavior, thereby enhancing detection accuracy and reducing errors.

In light of our findings, we propose the following recommendations for those engaged in the development or utilization of Android malware detection systems:
\begin{itemize}
    \item It is recommended that LLM integration be employed to enhance interpretability. It would be beneficial for organizations to consider incorporating LLMs such as ChatGPT into their existing detection pipeline. This not only enhances the efficacy of detection but also improves the interpretability of results, thereby increasing the system's reliability and facilitating audit processes.
    \item It is recommended that a hybrid detection model be employed. In order to achieve an optimal balance between decision speed and interpretability, a hybrid approach should be employed. The model will initially employ a traditional decision-based approach, while the LLM will provide an additional layer of analysis and insight.
    \item User-centered design is a priority for future detection systems. These systems should be user-friendly and interpretable, facilitating comprehension of the reasons behind detected threats among both technical and non-technical users. This will enhance adoption and trust in these systems.
\end{itemize}

\subsection{Why ChaGPT}
\noindent {We select ChatGPT as the LLM for the following reasons.

First, the goal of our research is not to compare or determine which LLM is superior in the field of Android malware detection and analysis. Instead, our focus is to demonstrate that LLMs have the potential to transform the existing detection paradigm. We aim to inform developers and users that malware detection methodologies in the LLM era should evolve. Rather than solely relying on traditional decision-oriented methods, greater emphasis should be placed on providing analysis and explanations to enhance understanding and usability.

Second,  This research began in 2023. At the time, Claude 1.0 was primarily designed as a chatbot with limited functionality. When Claude 2.0 was released, its performance fell short of ChatGPT 4 \cite{Borji2023BattleOT}. It was not until June 2024, with the release of Claude 3.5 Sonnet, that its performance surpassed ChatGPT 4 \footnote{\url{https://www.anthropic.com/news/claude-3-5-sonnet}}. However, Claude imposed strict usage restrictions, including requirements for billing and IP address consistency, as well as limitations on access in certain regions. Since our primary authors are based in Greater China, these restrictions made it impractical to utilize Claude in our research. In contrast, ChatGPT's more lenient access policies allowed us to integrate it seamlessly into our workflow.

Finally, due to hardware constraints and the lack of comprehensive datasets (not only in terms of sample size but also missing contextual information such as application background data, etc.), training a large-scale open-source LLM was not feasible. ChatGPT offered a ready-to-use solution with robust capabilities, enabling us to conduct our research effectively. That said, we are actively collecting relevant datasets to support the future development of an open-source LLM specifically designed for Android malware detection.

These considerations made ChatGPT the most convenient and accessible option for our study. Looking ahead, we plan to build upon this work and develop a specialized LLM for Android malware detection using open-source models.

\subsection{THREATS TO VALIDITY}
\label{sec_threats}

\noindent\textbf{Internal threats to validity.} We construct suitable prompts to allow ChatGPT to provide analysis and explanations, as well as to give a rating on the level of maliciousness. However, due to the uncertainty of ChatGPT, when analyzing the same application multiple times, the explanations and ratings given are not entirely consistent. Therefore, it is impossible to determine which answer is more truthful and effective. In the future, we hope to eliminate this uncertainty by constructing a large model specifically for Android malware detection.

{\setlength{\parindent}{0cm}
\textbf{External threats to validity.} First, our study explores the influence of a large model known as ChatGPT. Presently, numerous outstanding conversational large language models have yet to undergo similar investigations. Secondly, the dataset we employed offers both static and dynamic features; however, it has limitations in terms of scope, type, and the absence of contextual information, such as application usage. These limitations constrain ChatGPT's performance. In the future, we plan to gather more comprehensive datasets and investigate multiple large models.

\section{Related work}
\label{R-work}
\noindent \textbf{\textit{Android malware detection}}:  Android malware detection techniques can be categorized into two forms: static analysis and dynamic analysis. Static analysis involves decompiling the APK file to obtain the source code and metadata files, followed by extraction of significant features for malware detection and analysis such as the application programming interface \cite{r33,r34}, permissions \cite{r24,r35,r36,r37}, intent~\cite{r38}, function call graph~\cite{r39,r40,r41}, control flow~\cite{r42}, and grayscale~\cite{r43}. Li et al.~\cite{LiTosem2024} propose a meta-learning-based multi-classifier that uses multiple features such as permissions and APIs to classify malware and achieves good results. Although the meta-classifier learns meta-knowledge in the dataset, it cannot provide a basis for making decisions. Dynamic analysis~\cite{r44,r45,r46,r47,r48,r49} involves the extraction of features by monitoring the application's execution in real-time. Ficco~\cite{Ficco2022} combines general and specialized detectors to enhance the randomness of detection and improve the overall detection rate. Although this solution combines static and dynamic analysis, it is still decision-oriented and does not consider the interpretability of the solution. With the development of artificial intelligence, many related researches focus on models, such as SVM~\cite{r40,r50,r51}, and decision tree (DT)~\cite{r54}, RF~\cite{r24,r55}, AdaBoost~\cite{r43} and ensemble learning algorithm~\cite{r58,r59}, to classify malware. In addition, even deep learning methods~\cite{r26,r57} and meta-learning methods \cite{r64,r65} are directly used for classification. Although these methods have shown superior capabilities in detecting malware, they lack the ability to explain and analyze, and are unable to analyze the features of each input and inform developers and users of the functions and potential threats represented by these features.

\textbf{\textit{ChatGPT for SE}}: ChatGPT, with its powerful analysis and understanding capabilities, has been applied to various tasks in software engineering \cite{Emenike, mate, zhang}. Feng et al. \cite{Feng} utilize crowdsourced social data to study the code generation performance of ChatGPT. Sakib et al. \cite{sakib} explore the efficacy of ChatGPT in solving programming problems, examining the correctness and efficiency of its solutions in terms of time and memory complexity. Sun et al. \cite{sun} conduct research to compare and evaluate the capabilities of ChatGPT in automatic code summarization. These studies have solely concentrated on ChatGPT's capacity for generation, without delving into its capabilities for analysis and interpretation.

\section{Conclusion and Future Work}
\label{coclusion}
ChatGPT has demonstrated exceptional interpretation and analysis capabilities, making it applicable to a variety of tasks. We are the first to study the influence of ChatGPT on Android malware detection tasks. We select state-of-the-art Android malware detection solutions for comparison with ChatGPT. The research indicates that existing detection solutions suffer not only from dataset bias but also lack explanation and analysis capabilities. Given the existence of these problems and the excellent interpretability of ChatGPT, we identify a novel perspective for detecting Android malware, that is, less decision-making, and more explanation. In the future, we aim to enhance and expand the existing datasets by addressing gaps and completing missing information, such as application background details. Additionally, we plan to develop a specialized large language model tailored for Android malware detection, with a focus on improving both detection accuracy and interpretability.

\bibliographystyle{ACM-Reference-Format}
\bibliography{REF}

\end{document}